\documentclass[aps,twocolumn]{revtex4}
\usepackage{epsfig}
\newcommand{\be}{\begin{equation}}
\newcommand{\ee}{\end{equation}}
\newcommand{\bea}{\begin{eqnarray}}
\newcommand{\eea}{\end{eqnarray}}
\newcommand{\ba}{\begin{array}}
\newcommand{\ea}{\end{array}}

\begin{document}
\title{Estimating Mutual Information}
\author{Alexander Kraskov, Harald St\"ogbauer, and Peter Grassberger}
\affiliation{John-von-Neumann Institute for Computing, Forschungszentrum
J\"ulich, D-52425 J\"ulich, Germany}

\date{\today}
\begin{abstract}
We present two classes of improved estimators for mutual information
$M(X,Y)$, from samples of random points distributed according to some joint 
probability density $\mu(x,y)$. 
In contrast to conventional estimators based on binnings, they are based 
on entropy estimates from $k$-nearest neighbour distances. This means 
that they are data efficient (with $k=1$ we resolve structures down to 
the smallest possible scales), adaptive (the resolution is higher where data are 
more numerous), and have minimal bias. Indeed, the bias of the underlying 
entropy estimates is mainly due to non-uniformity of the density at the 
smallest resolved scale, giving typically systematic errors which scale as 
functions of $k/N$ for $N$ points.
Numerically, we find that both families become {\it exact} for independent
distributions, i.e. the estimator $\hat M(X,Y)$ vanishes (up to statistical
fluctuations) if $\mu(x,y) = \mu(x) \mu(y)$. This holds for all tested 
marginal distributions and for all dimensions of $x$ and $y$. In addition,
we give estimators for redundancies between more than 2 random variables.
We compare our algorithms in detail with existing algorithms.
Finally, we demonstrate the usefulness of our estimators for assessing 
the actual independence of components obtained from independent component 
analysis (ICA), for improving ICA, and for estimating the reliability of 
blind source separation.
\end{abstract}
\maketitle

\section{Introduction}

Among the measures of independence between random variables, mutual 
information (MI) is singled out by its information theoretic background 
\cite{cover-thomas}. In contrast to the linear 
correlation coefficient, it is sensitive also to dependencies which do not 
manifest themselves in the covariance. Indeed, MI is zero if and only if the 
two random variables are strictly independent. The latter is also true for 
quantities based on Renyi entropies \cite{renyi}, and these are often easier 
to estimate (in particular if their order is 2 or some other integer $> 2$). 
Nevertheless, MI is unique in its close ties to Shannon entropy and the 
theoretical advantages derived from this. Some well known properties of MI and 
some simple consequences thereof are collected in the appendix.

But it is also true that estimating MI is not always easy. Typically, one 
has a set of $N$ bivariate measurements, $z_i = (x_i,y_i), \, i=1,\ldots N$,
which are assumed to be iid (independent identically distributed) realizations
of a random variable $Z = (X,Y)$ with density $\mu(x,y)$. Here, $x$ and $y$ 
can be either scalars or can be elements of some higher dimensional space.
In the following we shall assume that the density is a proper smooth function,
although we could also allow more singular densities. All we need is that the 
integrals written below exist in some sense. In particular, we will always assume 
that $0 \log(0) =0$, i.e. we do not have to assume that densities are strictly
positive. The marginal densities of $X$ and $Y$ are $\mu_x(x) = \int dy \mu(x,y)$ 
and $\mu_y(y) = \int dx \mu(x,y)$. The MI is defined as
\be
   I(X,Y) = \int\!\!\!\int dx dy \;\mu(x,y) \;\log{\mu(x,y)\over \mu_x(x)\mu_y(y)}\;.
   \label{mi}
\ee
The base of the logarithm determines the units in which information is measured.
In particular, taking base two leads to information measured in bits. In 
the following, we always will use natural logarithms.
The aim is to estimate $I(X,Y)$ from the set $\{z_i\}$ alone, without knowing 
the densities $\mu, \mu_x$, and $\mu_y$.

One of the main fields where MI plays an important role, at least conceptually,
is independent component analysis (ICA) \cite{robe-ever,hyvar2001}. In the ICA 
literature, very 
crude approximations to MI based on cumulant expansions are popular because of 
their ease of use. But they are valid only for distributions close to Gaussians and
can mainly be used for ranking different distributions by interdependence, much 
less for estimating the actual dependences. Expressions obtained by entropy
maximalization using averages of some functions of the sample data as constraints
\cite{hyvar2001} are more robust, but are still very crude approximations.
Finally, estimates based on explicit parameterizations of the densities might be 
useful but are not very efficient. More promising are methods based on kernel 
density estimators \cite{moon95,steuer02}. We will not pursue these here either, 
but we will comment on them in Sec. IV.A.

The most straightforward and widespread approach for estimating MI more precisely 
consists in partitioning the supports of $X$ and $Y$ into bins of finite size, and 
approximating Eq.(\ref{mi}) by the finite sum 
\be
   I(X,Y) \approx I_{\rm binned}(X,Y) \equiv 
           \sum_{ij} p(i,j) \log{p(i,j) \over p_x(i) p_y(j)}\;,
   \label{miapprox}
\ee
where $p_x(i) = \int_i dx \;\mu_x(x)$, $p_y(j) = \int_j dy\; \mu_y(y)$, and 
$p(i,j) = \int_i \int_j dx dy \;\mu(x,y)$
-- and $\int_i$ means the integral over bin $i$.
An estimator of $I_{\rm binned}(X,Y)$ is obtained by counting the numbers of 
points falling into the various bins. If $n_x(i)\;(n_y(j))$ is the number of points 
falling into the $i$-th bin of $X$ ($j$-th bin of $Y$), and $n(i,j)$ is the number 
of points in their intersection, then we approximate $p_x(i) \approx n_x(i)/N$,
$p_y(j) \approx n_y(j)/N$, and $p(i,j) \approx n(i,j)/N$.
It is easily seen that the r.h.s. of Eq.(\ref{miapprox}) indeed converges to $I(X,Y)$
if we first let $N \to \infty$ and then let all bin sizes tend to zero,
if all densities exist as proper (not necessarily smooth) functions. If not, i.e.
if the distributions are e.g. (multi-)fractal, this convergence might no longer 
be true. In that case, Eq.(\ref{miapprox}) would define resolution dependent 
mutual entropies which diverge in the limit of infinite resolution. Although
the methods developed below could be adapted to apply also to that case, we 
shall not do this in the present paper.

The bin sizes used in Eq.(\ref{miapprox}) do not need to be the same for all bins.
Optimized estimators \cite{fraser-swinney,darbellay-vajda} use indeed adaptive
bin sizes which are essentially geared at having equal numbers $n(i,j)$ for all
pairs $(i,j)$ with non-zero measure. While such estimators are much better than
estimators using fixed bin sizes, they still have systematic errors which 
result on the one hand from approximating $I(X,Y)$ by $I_{\rm binned}(X,Y)$, 
and on the other hand by approximating (logarithms of) probabilities by 
(logarithms of) frequency ratios. The latter could be presumably minimized by
using corrections for finite $n_x(i)$ resp. $n(i,j)$ \cite{grass88}. These 
corrections are in the form of asymptotic series which diverge for finite $N$,
but whose first 2 terms improve the estimates in typical cases. The first 
correction term -- which often is not sufficient -- was taken into account
in \cite{roulston99,steuer02}.

In the present paper we will not follow these lines, but rather estimate MI
from $k$-nearest neighbour statistics. There exists an extensive literature
on such estimators for the simple Shannon entropy 
\be
   H(X) = -\int dx \mu(x) \log \mu(x),
\ee
dating back at least to \cite{dobrushin,vasicek}. But it 
seems that these methods have never been used for estimating MI. In 
\cite{vasicek,dude-meul,es,ebrahimi,correa,tsyb-meul,wiecz-grze} it is 
assumed that $x$ is one-dimensional, so that the $x_i$ can be ordered by
magnitude and $x_{i+1}-x_i \to 0$ for $N\to \infty$. In the simplest case, 
the estimator based only on these distances is 
\be 
   H(X) \approx {1\over N-1} \sum_{i=1}^{N-1} \log(x_{i+1}-x_i) + \psi(1) -
        \psi(N)\;.
   \label{vasi}
\ee
Here, $\psi(x)$ is the digamma function, $\psi(x) = \Gamma(x)^{-1} d\Gamma(x)/dx$.
It satisfies the recursion $\psi(x+1) = \psi(x)+1/x$ and $\psi(1) = -C$ where 
$C = 0.5772156\ldots$ is the Euler-Mascheroni constant. For large $x$, 
$\psi(x) \approx \log x -1/2x$. Similar formulas exist which use $x_{i+k}-x_i$
instead of $x_{i+1}-x_i$, for any integer $k<N$.

Although Eq.(\ref{vasi}) and its generalizations to $k>1$ seem to give the 
best estimators of $H(X)$, they cannot be used for MI because it is not 
obvious how to generalize them to higher dimensions. Here we have to use a 
slightly different approach, due to \cite{koza-leon} (see also 
\cite{grass85,somorjai86}; the latter authors were only interested in 
fractal measures and estimating their information dimensions, but the basic 
concepts are the same as in estimating $H(X)$ for smooth densities). 

Assume some metrics to be given on the spaces spanned by $X, Y$ and $Z=(X,Y)$. 
We can then rank, for each point $z_i = (x_i,y_i)$, its neighbours by distance
$d_{i,j}= || z_i-z_j||$: $d_{i,j_1} \leq d_{i,j_2} \leq d_{i,j_3} \leq \ldots$.
Similar rankings can be done in the subspaces $X$ and $Y$. The basic idea of
\cite{koza-leon,grass85,somorjai86} is to estimate $H(X)$ from the average
distance to the $k$-nearest neighbour, averaged over all $x_i$. Details will
be given in Sec.II. Mutual information could be obtained by estimating in this
way $H(X)$, $H(Y)$ and $H(X,Y)$ separately and using \cite{cover-thomas}
\be
   I(X,Y) = H(X)+H(Y)-H(X,Y)\;.
\ee
But this would mean that the errors made in the individual estimates would 
presumably not cancel, and therefore we proceed differently.

Indeed we will present two slightly different algorithms, both based on 
the above idea. Both use for the space $Z=(X,Y)$ the maximum norm,
\be
   ||z-z'|| = \max\{||x-x'||,||y-y'||\},
\ee
while any norms can be used for $||x-x'||$ and $||y-y'||$ (they need not be 
the same, as these spaces could be completely different). Let us denote 
by $\epsilon(i)/2$ the distance from $z_i$ to its $k$-th neighbour, and by 
$\epsilon_x(i)/2$ and $\epsilon_y(i)/2$ the distances between the same points 
projected into the $X$ and 
$Y$ subspaces. Obviously, $\epsilon(i) = \max\{\epsilon_x(i),\epsilon_y(i)\}$.

In the first algorithm, we count the number $n_x(i)$ of points $x_j$ whose
distance from $x_i$ is strictly less than $\epsilon(i)/2$, and similarly 
for $y$ instead of $x$. This is illustrated in Fig.~1a. Notice that 
$\epsilon(i)$ is a random (fluctuating)
variable, and therefore also $n_x(i)$ and $n_y(i)$ fluctuate. We denote 
by $\langle \ldots \rangle$ averages both over all $i\in [1,\ldots N]$ and 
over all realizations of the random samples,
\be
   \langle \ldots \rangle = N^{-1} \sum_{i=1}^N {\sf E}[\ldots(i)]\;.
\ee
The estimate for MI is then
\be
   I^{(1)}(X,Y) = \psi(k) - \langle \psi(n_x+1) +\psi(n_y+1) \rangle + \psi(N).
   \label{i1}
\ee

\begin{figure}
  \begin{center}
    \psfig{file=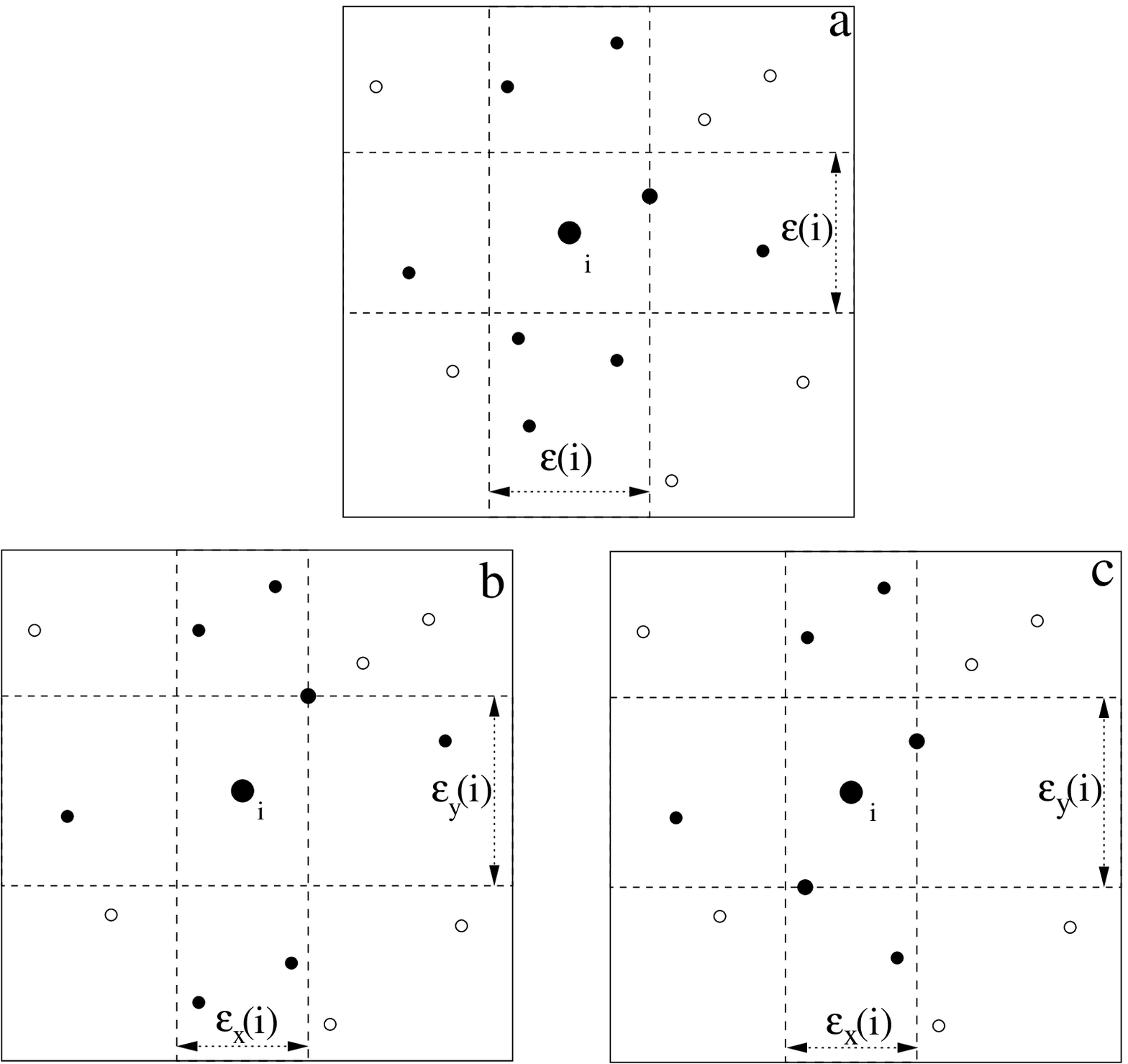,width=7.6cm,angle=0}
 \caption{Panel (a): Determination of $\epsilon(i)$, $n_x(i)$ and $n_y(i)$ in the 
    first algorithm, for $k=1$ and some fixed $i$. In this example, $n_x(i)=5$ and
    $n_y(i)=3$.\\
    Panels (b),(c): Determination of $\epsilon_x(i)$, $\epsilon_y(i)$, 
    $n_x(i)$ and $n_y(i)$ in the second algorithm. Panel (b) shows a case where 
    $\epsilon_x(i)$ and $\epsilon_y(i)$ are determined by the same point, while 
    panel (c) shows a case where they are determined by different points.}
 \label{algo}
 \end{center}
 \end{figure}

Alternatively, in the second algorithm, we replace $n_x(i)$ and $n_y(i)$ by 
the number of points with $||x_i-x_j|| \leq \epsilon_x(i)/2$ and 
$||y_i-y_j|| \leq \epsilon_y(i)/2$ (see Figs.~1b and 1c). The estimate for MI
is then 
\be
   I^{(2)}(X,Y) = \psi(k) - 1/k - \langle \psi(n_x)+ \psi(n_y) \rangle + \psi(N).
   \label{i2}
\ee
The derivations of Eqs.(\ref{i1}) and (\ref{i2}) will be given in Sec.II.
There we will also give formulas for generalized redundancies in higher
dimensions,
\bea
   I(X_1,X_2, \ldots X_m)& = &H(X_1)+H(X_2)+\ldots +H(X_m) \nonumber \\ 
            &-&H(X_1,X_2, \ldots X_m).
   \label{redund}
\eea
In general, both formulas give very similar results. For the same $k$, Eq.(\ref{i1})
gives slightly smaller statistical errors (because $n_x(i)$ and $n_y(i)$ tend
to be larger and have smaller relative fluctuations), but have larger 
systematic errors. 
The latter is only severe if we are interested in very high dimensions where 
$\epsilon(i)$ tends typically to be much larger than the marginal $\epsilon_{x_j}(i)$.
In that case the second algorithm seems preferable. Otherwise, both can be 
used equally well.

A systematic study of the performance of Eqs.(\ref{i1}) and (\ref{i2}) and 
comparison with previous algorithms will be given in Sec.III. Here we will 
just show results of $I^{(2)}(X,Y)$ for Gaussian distributions. Let $X$ and $Y$
be Gaussians with zero mean and unit variance, and with covariance $r$. In this 
case $I(X,Y)$ is known exactly \cite{darbellay-vajda},
\be
   I_{\rm Gauss}(X,Y) = -{1\over 2} \log(1-r^2)\;.
   \label{gauss-MI}
\ee
In Fig.~2 we show the errors $I^{(2)}(X,Y) - I_{\rm Gauss}(X,Y)$ for various 
values of $r$, obtained from a large number (typically $10^5-10^7$) of realizations.
We show only results for $k=1$, plotted against $1/N$. Results for $k>1$ are 
similar. To a first approximation $I^{(1)}(X,Y)$ and $I^{(2)}(X,Y)$ depend 
only on the ratio $k/N$.

The most conspicuous feature seen in Fig.~2, apart from the fact that indeed
$I^{(2)}(X,Y) - I_{\rm Gauss}(X,Y) \to 0$ for $N\to \infty$, is that the 
systematic error is compatible with zero for $r=0$, i.e. when the two Gaussians
are uncorrelated. We checked this with high statistics runs for many different 
values of $k$ and $N$ (a priori one should expect that systematic errors 
become large for very small $N$), and for many more distributions (exponential,
uniform, ...). In all cases we found that both $I^{(1)}(X,Y)$ and $I^{(2)}(X,Y)$ 
become exact for independent variables. Moreover, the same seems to be true 
for higher order redundancies. We thus have the 

{\bf Conjecture:} Eqs.(\ref{i1}) and (\ref{i2}) are exact for independent $X$ and 
$Y$, i.e. $I^{(1)}(X,Y) = I^{(2)}(X,Y) = 0$ if and only if $I(X,Y)=0$.

We have no proof for this very surprising result. We have numerical indications 
that moreover
\be
   {|I^{(1,2)}(X,Y) - I(X,Y)|\over I(X,Y)} \leq {\rm const} 
\ee
as $X$ and $Y$ become more and more independent, but this is much less clean 
and therefore much less sure.

In Sec.II we shall give formal arguments for our estimators, and for generalizations
to higher dimensions. Detailed numerical results for cases where the exact MI is 
known will be given in Sec.III. In Sec.IV.A we give two preliminary applications to 
gene expression data and to ICA.
Conclusions are drawn in the last section, Sec.V. Finally, some general aspects
of MI are recalled in an appendix.

\begin{figure}
  \begin{center}
    \psfig{file=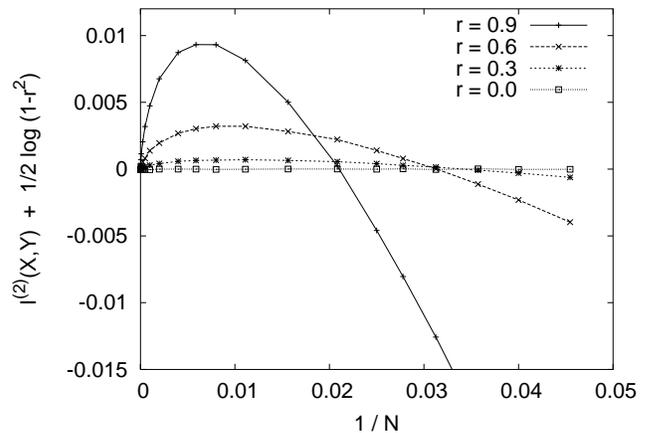,width=6.0cm,angle=270}
 \caption{Estimates of $I^{(2)}(X,Y) - I_{\rm exact}(X,Y)$ for Gaussians with
   unit variance and covariances $r=0.9, 0.6, 0.3$, and 0.0 (from top to bottom),
   plotted against $1/N$. In all cases $k=1$. The number of realizations is 
   $>2\times 10^6$ for $N<=1000$, and decreases to $\approx 10^5$ for $N=40,000$.
   Error bars are smaller than the sizes of the symbols.}
 \label{gauss}
 \end{center}
 \end{figure}

\section{Formal Developments}

\subsection{Kozachenko - Leonenko Estimate for Shannon Entropies}

We first review the derivation of the Shannon entropy estimate 
\cite{grass85,somorjai86,koza-leon,victor}, since
the estimators for MI are obtained by very similar arguments.

Let $X$ be a continuous random variable with values in some metric
space, i.e. there is a distance function $||x-x'||$ between any 
two realizations of $X$, and let the density $\mu(x)$ exist as a 
proper function. Shannon entropy is defined as 
\be
   H(X) = -\int dx \mu(x) \log \mu(x)\;,     \label{s1}
\ee
where ``log" will always mean natural logarithm so that information
is measured in natural units. Our aim is to estimate $H(X)$ from a 
random sample $(x_1\ldots x_N)$ of $N$ realizations of $X$.

The first step is to realize that Eq.(\ref{s1}) can be understood 
(up to the minus sign) as an average of $\log \mu(x)$. If we had 
unbiased estimators $\widehat{\log \mu(x)}$ of the latter, we would have 
an unbiased estimator
\be
   \hat{H}(X) = - N^{-1} \sum_{i=1}^N \widehat{\log \mu(x_i)}\;.
   \label{H_muav}
\ee

In order to obtain the estimate $\widehat{\log \mu(x_i)}$, we consider
the probability distribution $P_k(\epsilon)$ for the distance between 
$x_i$ and its $k$-th nearest neighbour. The probability $P_k(\epsilon)
d\epsilon$ is equal to the chance that there is one point within
distance $r\in [\epsilon/2,\epsilon/2+d\epsilon/2]$ from $x_i$, that there
are $k-1$ other points at smaller distances, and that $N-k-1$ points
have larger distances from $x_k$. Let us denote by $p_i$ the mass of 
the $\epsilon$-ball centered at $x_i$, $p_i(\epsilon) = 
\int_{||\xi-x_i||<\epsilon/2} d\xi \mu(\xi)$. Using the trinomial
formula we obtain
\bea
   P_k(\epsilon)d\epsilon & = & {(N-1)!\over 1!(k-1)!(N-k-1)!}\; \times \nonumber \\
     &\times& {dp_i(\epsilon)\over d\epsilon} d\epsilon\;\times\
     p_i^{k-1} \times (1-p_i)^{N-k-1}
\eea
or 
\be
   P_k(\epsilon) = k {N-1\choose k} \;
     {dp_i(\epsilon)\over d\epsilon} \;p_i^{k-1} (1-p_i)^{N-k-1} \;.
   \label{Pk}
\ee

One easily checks that this is correctly normalized, $\int d\epsilon
P_k(\epsilon) = 1$. Similarly, one can compute the expectation value
of $\log p_i(\epsilon)$
\bea
   {\sf E}(\log p_i)
      & = &\int_0^\infty d\epsilon\; P_k(\epsilon) \log p_i(\epsilon)  \nonumber \\
      & = & k {N-1\choose k} \; \int_0^1 dp\; p^{k-1}(1-p)^{N-k-1}
      \log p \nonumber \\
      & = &\psi(k) - \psi(N)\;,
      \label{logp}
\eea
where $\psi(x)$ is the digamma function. The expectation is taken here
over the positions of all other $N-1$ points, with $x_i$ kept fixed.
An estimator for $\log \mu(x)$ is then obtained by assuming that $\mu(x)$
is constant in the entire $\epsilon$-ball. The latter gives
\be
   p_i(\epsilon) \approx c_d \epsilon^d \mu(x_i) \;.
      \label{p_approx}
\ee
where $d$ is the dimension of $x$ and $c_d$ is the volume of the $d$-dimensional 
unit ball. For the maximum norm one has simply $c_d =1$, while 
$c_d = \pi^{d/2}/\Gamma(1+d/2)/2^d$ for Euclidean norm.

Using Eqs.(\ref{logp}) and (\ref{p_approx}) one obtains
\be
   \log \mu(x_i) \approx \psi(k) - \psi(N) - d\; {\sf E}(\log\epsilon) - \log c_d \;,
\ee
which finally leads to
\be
   \hat{H}(X) = - \psi(k) + \psi(N) + \log c_d + {d\over N} \sum_{i=1}^N \log\epsilon(i)
   \label{KL}
\ee
where $\epsilon(i)$ is twice the distance from $x_i$ to its $k$-th neighbour.
      
From the derivation it is obvious that Eq.(\ref{KL}) would be unbiased,
if the density $\mu(x)$ were strictly constant. The only 
approximation is in Eq.(\ref{p_approx}). For points on a torus (e.g. when $x$ 
is a phase) with a strictly positive density one can easily estimate the 
leading corrections to Eq.(\ref{p_approx}) for large $N$. One finds that they
are $O(1/N^2)$ and that they scale, for large $k$ and $N$, as $\sim (k/N)^2$.
In most other cases (including e.g. Gaussians and uniform densities in bounded
domains with a sharp cut-off) it seems numerically that the error is 
$\sim k/N$ or $\sim k/N \log(N/k)$. 

\subsection{Mutual Informations: Estimator $I^{(1)}(X,Y)$}

Let us now consider the joint random variable $Z=(X,Y)$ with maximum norm.
Again we take one of the $N$ points $z_i$ and consider the distance $\epsilon/2$ 
to its $k$-th neighbour. Again this is a random variable with distribution
given by Eq.(\ref{Pk}). Also Eq.(\ref{logp}) holds without changes. The first 
difference to the previous subsection is in Eq.(\ref{p_approx}), where we 
have to replace $d$ by $d_Z = d_X+d_Y$, $c_d$ by $c_{d_X}c_{d_Y}$, and of 
course $x_i$ by $z_i=(x_i,y_i)$. With these modifications we obtain therefore
\bea
   \hat{H}(X,Y) &=& \psi(k) - \psi(N) - \log(c_{d_X}c_{d_Y}) \nonumber \\
     &-& {d_X+d_Y\over N} \sum_{i=1}^N \log\epsilon(i)\;.
\eea

In order to obtain $I(X,Y)$, we have to subtract this from estimates for 
$H(X)$ and $H(Y)$. For the latter, we could use Eq.(\ref{KL}) directly with
the same $k$. But this would mean that we would effectively use different 
distance scales in the joint and marginal spaces. For any fixed $k$, the 
distance to the $k$-th neighbour in the joint space will be larger than 
the distances to the neighbours in the marginal spaces. Since the bias in 
Eq.(\ref{KL}) from the non-uniformity of the density depends of course 
on these distances, the biases in $\hat{H}(X)$, $\hat{H}(Y)$, and in
$\hat{H}(X,Y)$ would not cancel. 

To avoid this, we notice that Eq.(\ref{KL}) holds for {\it any} value 
of $k$, and that we do not have to choose a fixed $k$ when estimating 
the marginal entropies. Assume, as in Fig.~1a, that the $k$-th neighbour of 
$x_i$ is on one of the vertical sides of the square of size $\epsilon(i)$. 
In this case, if there are altogether $n_x(i)$ points within the vertical
lines $x = x_i\pm \epsilon(i)/2$, then $\epsilon(i)/2$ is the distance to the 
$(n_x(i)+1)-$st neighbour of $x_i$, and 
\bea
   \hat{H}(X) & = & {1\over N} \sum_{i=1}^N \psi(n_x(i)+1) - \psi(N) \nonumber \\
     & - & \log c_{d_X} - {d_X\over N} \sum_{i=1}^N \log\epsilon(i)\;.
   \label{HX}
\eea
For the other direction (the $y$ direction in Fig.~1a) this is not exactly true,
i.e. $\epsilon(i)$ is not exactly equal to twice the distance to the $(n_y(i)+1)-$st
neighbour, if $n_y(i)$ is analogously defined as the number of points with 
$||y_j-y_i|| <\epsilon(i)/2$. Nevertheless we can consider Eq.(\ref{HX}) also
as a good approximation for $H(Y)$, if we replace everywhere $X$ by $Y$ in
its right hand side (this approximation becomes exact when $n_y(i)\to\infty$, 
and thus also when $N\to\infty$). If we do this, subtracting $\hat{H}(X,Y)$ 
from $\hat{H}(X)+\hat{H}(Y)$ leads directly to Eq.(\ref{i1}).

These arguments can be easily extended to $m$ random variables and lead to
\bea
   I^{(1)}(X_1,X_2, \ldots X_m)& = & \psi(k) + (m-1)\psi(N)  \\
    & - & \langle \psi(n_{x_1}) + \psi(n_{x_2}) + \ldots \psi(n_{x_m}) \rangle\;.
             \nonumber
\eea

\subsection{Mutual Informations: Estimator $I^{(2)}(X,Y)$}

The main drawback of the above derivation is that the Kozachenko-Leonenko
estimator is used correctly in only one marginal direction. This seems
unavoidable if one wants to stick to ``balls", i.e. to (hyper-)cubes in 
the joint space. In order to avoid it we have to switch to (hyper-)rectangles.

Let us first discuss the case of two marginal variables $X$ and $Y$, and 
generalize later to $m$ variables $X_1, \ldots X_m$. As illustrated in Figs.~1b
and 1c, there are two cases to be distinguished (all other cases, where more
points fall onto the boundaries $x_i\pm\epsilon_x(i)/2$ and 
$y_i\pm\epsilon_y(i)/2$, have zero probability; see however the third 
paragraph of Sec.III): Either the two sides $\epsilon_x(i)$ and $\epsilon_y(i)$
are determined by the same point (Fig.~1b), or by different points (Fig.~1c).
In either case we have to replace $P_k(\epsilon)$ by a 2-dimensional 
density,  
\be
   P_k(\epsilon_x,\epsilon_y) = P_k^{(b)}(\epsilon_x,\epsilon_y)+
              P_k^{(c)}(\epsilon_x,\epsilon_y)
\ee
with 
\be
   P_k^{(b)}(\epsilon_x,\epsilon_y) 
    = {N-1\choose k} \;
     {d^2[q_i^k]\over d\epsilon_x d\epsilon_y} \; (1-p_i)^{N-k-1}
   \label{Pka}
\ee
and 
\be
   P_k^{(c)}(\epsilon_x,\epsilon_y) 
    = (k-1) {N-1\choose k} \;
     {d^2[q_i^k]\over d\epsilon_x d\epsilon_y} \; (1-p_i)^{N-k-1} \;.
   \label{Pkb}
\ee
Here, $q_i\equiv q_i(\epsilon_x,\epsilon_y)$ is the mass of the rectangle of 
size $\epsilon_x\times \epsilon_y$ centered at $(x_i,y_i)$, and $p_i$ is 
as before the mass of the square of size $\epsilon = \max\{\epsilon_x,\epsilon_y\}$.
The latter is needed since by using the maximum norm we guarantee that there are 
no points in this square which are not inside the rectangle.

Again we verify straightforwardly that $P_k$ is normalized, while we have now 
instead of Eq.(\ref{logp})
\bea
   {\sf E}(\log q_i)
      & = &\int\!\!\!\!\int_0^\infty d\epsilon_x d\epsilon_y 
          P_k(\epsilon_x,\epsilon_y) \log q_i(\epsilon_x,\epsilon_y)  \nonumber \\
      & = &\psi(k) -1/k - \psi(N)\;.
      \label{logp2}
\eea
Denoting now by $n_x(i)$ and $n_y(i)$ the number of points with distance
less {\it or equal} to $\epsilon_x(i)/2$ resp. $\epsilon_y(i)/2$, we arrive at
Eq.(\ref{i2}).

For the generalization to $m$ variables we have to consider $m$-dimensional 
densities $P_k(\epsilon_{x_1},\ldots \epsilon_{x_m})$. The number of 
distinct cases (analogous to the two cases shown in Figs.~1b and 1c) proliferates
as $m$ grows, but fortunately we do not have to consider all these cases 
explicitly. One sees easily that each of them contributes to $P_k$ a term
\be
   \propto {d^m[q_i^k]\over d\epsilon_{x_1}\ldots d\epsilon_{x_m}} \; (1-p_i)^{N-k-1}
\ee
The direct calculation of the proportionality factors would be extremely 
tedious (we did it for $m=3$), but it can be avoided by simply demanding that
the sum is correctly normalized. This gives
\bea
   P_k(\epsilon_{x_1},\ldots \epsilon_{x_m}) &=& k^{m-1} {N-1\choose k} 
     {d^m[q_i^k]\over d\epsilon_{x_1}\ldots d\epsilon_{x_m}}  \times\; \nonumber \\
    & \times &(1-p_i)^{N-k-1}\;.
\eea
Calculating again ${\sf E}(\log q_i) = \psi(k) -(m-1)/k - \psi(N)$ analytically and
approximating the density by a constant inside the hyper-rectangle, we 
obtain finally
\bea
   I^{(2)}(X_1,X_2, \ldots X_m)& = & \psi(k) -(m-1)/k \nonumber \\
    & + & (m-1)\psi(N)  \\
    & - & \langle \psi(n_{x_1}) + \psi(n_{x_2}) + \ldots \psi(n_{x_m}) \rangle\;.
             \nonumber
    \label{rectangl}
\eea

Before leaving this section, we should mention that we slightly cheated in 
deriving $I^{(2)}(X,Y)$ (and its generalization to $m>2$). Assume that in a
particular realization we have $\epsilon_x(i) < \epsilon_y(i)$, as in Fig.~1b,c.
In that case we know that there cannot be any point in the two rectangles
$[x_i-\epsilon_y(i)/2,x_i-\epsilon_x(i)/2]\times [y_i-\epsilon_y(i)/2,y_i+\epsilon_y(i)/2]$
and $[x_i+\epsilon_x(i)/2,x_i+\epsilon_y(i)/2]\times 
[y_i-\epsilon_y(i)/2,y_i+\epsilon_y(i)/2]$ (see Fig.~3).
While we have taken this correctly into account when estimating $H(X,Y)$ (where
it was crucial), we have neglected it in $H(X)$ and $H(Y)$. There, the corrections 
are $O(1/n_x)$ and $O(1/n_y)$, and should vanish for $N\to\infty$. It could be 
that their net effect vanishes, because they contribute with opposite signs to 
$H(X)$ and $H(Y)$. But we have no proof for it. Anyhow, due to the approximation
of constant density within each rectangle we cannot expect our estimates to be 
exact for finite $N$, and any justification ultimately relies on numerics.

\begin{figure}
  \begin{center}
\parbox{.27\textwidth}{\includegraphics[width=.24\textwidth,angle=0]{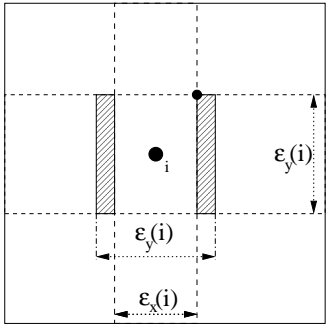}}
\parbox{.20\textwidth}{\caption[]{There cannot be any points inside the shaded rectangles. For method 2, 
   this means that the estimates of the marginal entropy $H(X)$ ($H(Y)$) 
   should be modified, since part of the area outside (inside) the stripe
   of with $\epsilon_x$ ($\epsilon_y$) is forbidden. This is neglected in 
   Eq.(\ref{i2}).}}
 \label{corr}
\end{center}
\end{figure}

\section{Implementation and Results}

\subsection{Some Implementation Details}

Mutual information is invariant under reparametrization of the marginal variables.
If $X' = F(X)$ and $Y' = G(Y)$ are homeomorphisms, then $I(X,Y) = I(X',Y')$ (see
appendix). This is in contrast to $H(X)$ which changes in general under a 
homeomorphism. This can be used to rescale both variables first to unit variance.
In addition, if the distributions are very skewed and/or 
rough, it might be a good idea to transform them such as to become more uniform (or 
at least single-humped and more or less symmetric). Although this is not required,
strictly spoken, it will in general reduce errors. One example is the gamma-exponential
distribution in 2 variables, $\mu(x,y)=x^\theta \exp(-x-xy)/\Gamma(\theta)$
for $x,y>0$ \cite{darb-vajd2000}, when $\theta < 1$. For $\theta \to 0$, the marginal
distributions develop $1/x$ resp. $1/y$ singularities (for $x\to 0$ and for 
$y\to \infty$, respectively), and the joint distribution is non-zero only in 
a very narrow region near the two axes. In this case our algorithm failed 
when applied directly, but it gave excellent results after transforming the 
variables to $x'=\log x$ and $y'=\log y$.

When implemented straightforwardly, the algorithm spends most of the CPU time 
for searching neighbours. In the most naive version, we need two nested loops 
through all points which gives a CPU time $O(N^2)$. While this is acceptable for 
very small data sets (say $N\le 300$), fast neighbour search algorithms are 
needed when dealing with larger sets. Let us assume that $X$ and $Y$ are scalars.
An algorithm with complexity $O(N\sqrt{k\;N})$ is then obtained by first ranking the
$x_i$ by magnitude (this can be done by any sorting algorithm such as quicksort),
and co-ranking the $y_i$ with them \cite{numrec}. Nearest neighbours of $(x_i,y_i)$
can then be obtained by searching $x$-neighbours on both sides of $x_i$ and 
verifying that their distance in $y$ direction is not too large. Neighbours in
the marginal subspaces are found even easier by ranking both $x_i$ and $y_i$.
Most results in this paper were obtained by this method which is suitable for 
$N$ up to a few thousands. The fastest (but also most complex) algorithm is 
obtained by using grids (`boxes') \cite{grass90,tisean}. Indeed, we use three grids:
A 2-dimensional one with box size $O(\sqrt{k/N})$ and two 1-dimensional ones
with box sizes $O(1/N)$. First the $k$ neighbours in 2-d space are searched using 
the 2-d grid, then the boxes at distances $\pm \epsilon$ from the central point 
are searched in the 1-d grids to find $n_x$ and $n_y$. If the distributions 
are smooth, this leads to complexity $O(\sqrt{k}N)$. The last algorithm is 
comparable in speed to the algorithm of \cite{darbellay-vajda}. For all three 
versions of our algorithm it costs only little additional CPU time if one evaluates, 
together with $I(X,Y)$ for some $k>1$, also the estimators for smaller $k$.

Empirical data usually are obtained with few (e.g. 12 or 16) binary digits, 
which means that many points in a large set may have identical coordinates.
In that case the numbers $n_x(i)$ and $n_y(i)$ need no longer be unique (the 
assumption of continuously distributed points is violated). If no precautions 
are taken, any code based on nearest neighbour counting is then bound to 
give wrong results. The simplest way out of this dilemma is to add very low
amplitude noise to the data ($\approx 10^{-10}$, say, when working with 
double precision) which breaks this degeneracy. We found this to give 
satisfactory results in all cases. 

Often, MI is estimated after {\it rank ordering} the data, i.e. after 
replacing the coordinate $x_i$ by the rank of the $i$-th point when sorted
by magnitude. This is equivalent to applying a monotonic transformation 
$x\to x',\; y\to y'$ to each coordinate which leads to a strictly uniform 
empirical density, $\mu'_x(x') = \mu'_y(x') = (1/N)\sum_{i=1}^N \delta(x'-i)$.
For $N\to\infty$ and $k \gg 1$ this
clearly leaves the MI estimate invariant. But it is not obvious that it 
leaves invariant also the estimates for finite $k$, since the transformation
is not smooth at the smallest length scale. We found numerically that 
rank ordering gives correct estimates also for small $k$, if the distance 
degeneracies implied by it are broken by adding low amplitude noise as 
discussed above. In particular, both estimators still gave zero MI for 
independent pairs. Although rank ordering can reduce statistical errors, 
we did not apply it in the following tests, and we did not study in detail
the properties of the resulting estimators.

\subsection{Results: Two-Dimensional Distributions}

We shall first discuss applications of our estimators to correlated 
Gaussians, mainly because we can in this way most easily compare with 
analytic results and with previous numerical analyses. In all cases we 
shall deal with Gaussians of unit variance and zero mean. For $m$ such 
Gaussians with covariance matrix $\sigma_{ik}\;i,k = 1\ldots m$, one 
has 
\be
   I(X_1,\ldots X_m) = -{1\over 2} \log({\rm det}(\sigma))\;.
  \label{gauss-analytic}
\ee
For $m=2$ and using the notation $r=\sigma_{XY}$, this gives 
Eq.(\ref{gauss-MI}).

\begin{figure}
  \begin{center}
    \psfig{file=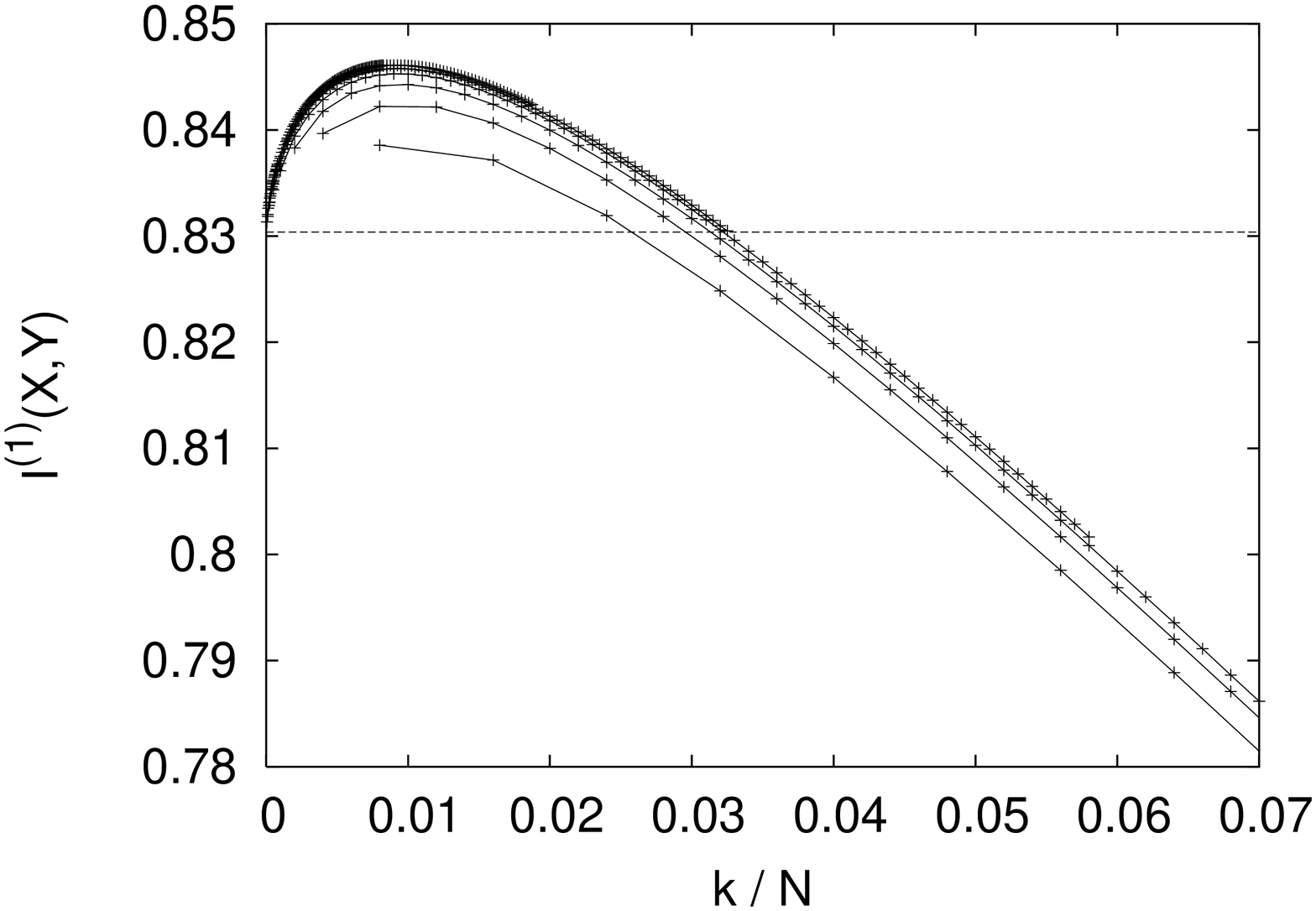,width=2.9cm,angle=270}
    \psfig{file=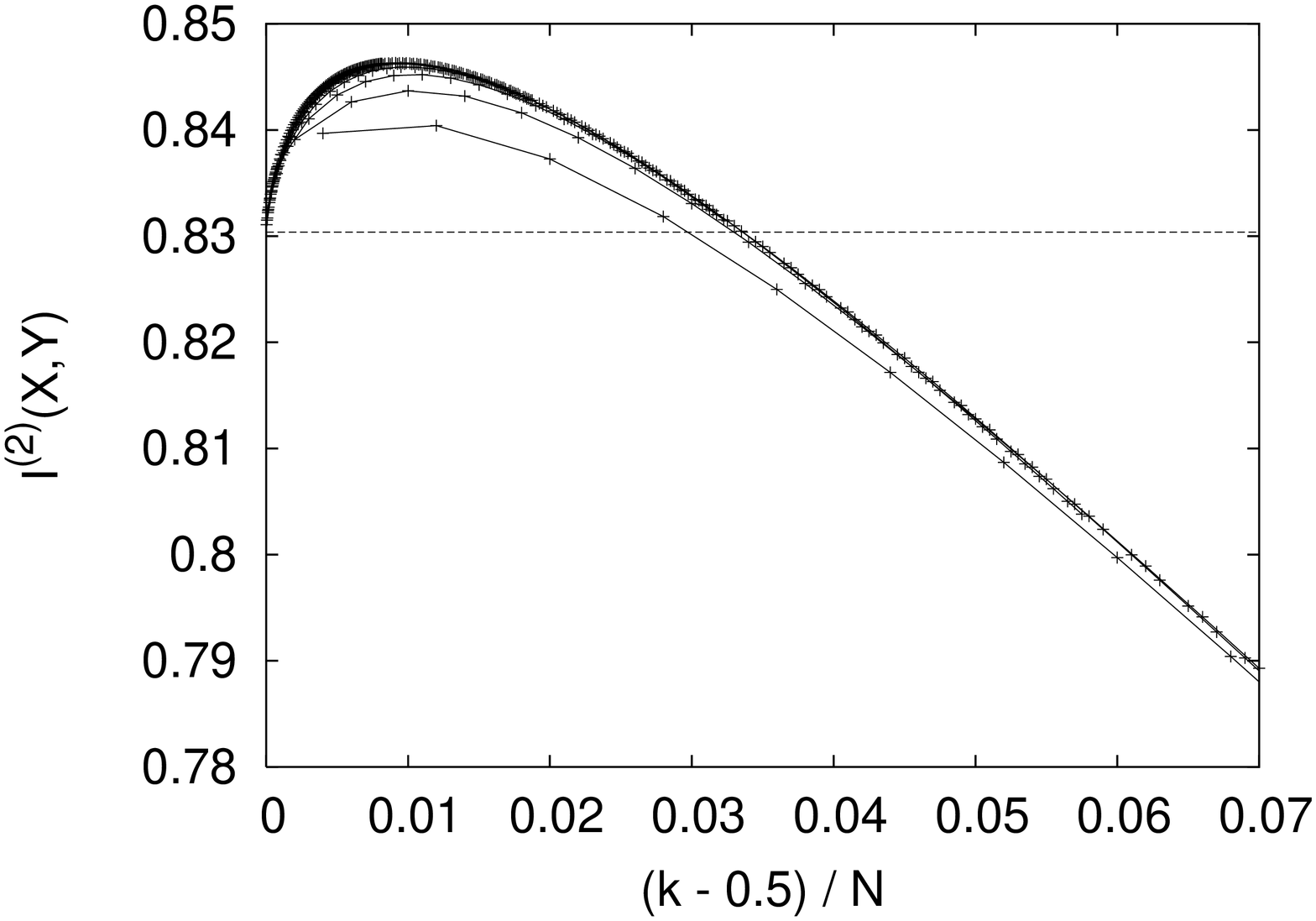,width=2.9cm,angle=270}
 \caption{Mutual information estimates $I^{(1)}(X,Y)$ (left panel) and 
    $I^{(2)}(X,Y)$ (right panel) for Gaussian deviates with unit 
    variance and covariance $r=0.9$, plotted against $k/N$ (left panel)
    resp. $(k-1/2)/N$ (right panel). Each curve 
    corresponds to a fixed value of $N$, with $N=125, 250, 500, 1000, 2000, 
    4000, 10000$ and $20000$, from bottom to top. Error bars are smaller 
    than the size of the symbols. The dashed line
    indicates the exact value $I(X,Y)= 0.830366$.}
 \label{gauss-kN}
\end{center}
\end{figure}

First results for $I^{(2)}(X,Y)$ with $k=1$ were already shown in Fig.~2. 
Results obtained with $I^{(1)}(X,Y)$ are very similar and would indeed
be hard to distinguish on this figure. In 
Fig.~\ref{gauss-kN} we compare values of $I^{(1)}(X,Y)$ (left panel) 
with those for $I^{(2)}(X,Y)$ (right panel) for different values of $N$ and 
for $r=0.9$. The horizontal axes show $k/N$ (left) and $(k-1/2)/N$ (right).
Except for very small values of $k$ and $N$, we observe scaling of the form 
\be
   I^{(1)}(X,Y) \approx \Phi({k\over N})\;,\quad 
             I^{(2)}(X,Y) \approx \Phi({k-1/2\over N})\;.
\ee
This is a general result and is found also for other distributions.
The scaling with $k/N$ of $I^{(1)}(X,Y)$ results simply from the fact that 
the number of neighbours within a fixed distance would scale $\propto N$,
if there were no statistical fluctuations. For large $k$ these fluctuations
should become irrelevant, and thus the MI estimate should depend only on 
the ratio $k/N$. For $I^{(2)}(X,Y)$ this argument has to be slightly 
modified, since the smaller one of $\epsilon_x$ and $\epsilon_y$ is determined
(for large $k$ where the situation illustrated in Fig.~1c dominates
over that in Fig.~1b) by $k-1$ instead of $k$ neighbours.

\begin{figure}
  \begin{center}
    \psfig{file=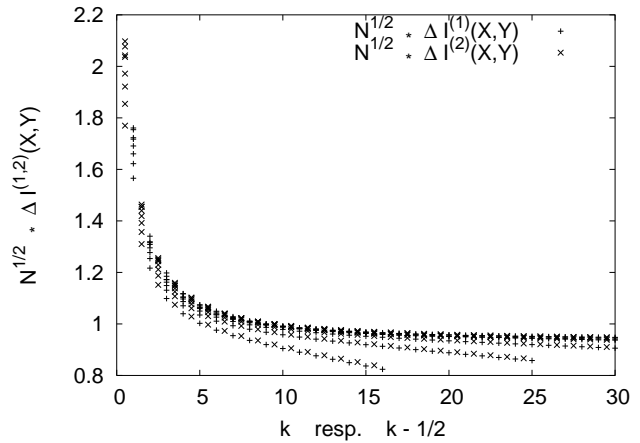,width=6cm,angle=270}
 \caption{Standard deviations of the estimates $I^{(1)}(X,Y)$ ($+$) 
    and $I^{(2)}(X,Y)$ ($\times$) for Gaussian deviates with unit
    variance and covariance $r=0.9$, multiplied by $\sqrt{N}$ and plotted 
    against $k$ ($I^{(1)}(X,Y)$) resp. $k-1/2$ ($I^{(2)}(X,Y)$). Each curve
    corresponds to a fixed value of $N$, with $N=125, 250, 500, 1000, 2000,
    4000, 10000$ and $20000$, from bottom to top. }
 \label{gauss-standdev}
\end{center}
\end{figure}

The fact that $I^{(2)}(X,Y)$ for a given value of $k$ is between 
$I^{(1)}(X,Y)$ for $k-1$ and $I^{(1)}(X,Y)$ for $k$ is also seen from the 
variances of the estimates. In Fig.~\ref{gauss-standdev} we show the 
standard deviations, again for covariance $r=0.9$. These statistical
errors depend only weakly on $r$. For $r=0$ they are roughly 10\% smaller.
As seen from Fig.~\ref{gauss-standdev}, the errors of $I^{(2)}(X,Y;k)$ 
are roughly half-way between those of $I^{(1)}(X,Y;k-1)$ and 
$I^{(1)}(X,Y;k)$. They scale roughly as $\sim \sqrt{N}$, except for 
very large $k/N$. Their dependence on $k$ does not follow a simple 
scaling law. The fact that statistical errors increase when $k$ decreases
is intuitively obvious, since then the width of the distribution of 
$\epsilon$ increases too. Qualitatively the same dependence of the 
errors was observed also for different distributions. For practical
applications, it means that one should use $k>1$ in order to reduce
statistical errors, but too large values of $k$ should be avoided 
since then the increase of systematic errors outweighs the decrease
of statistical ones. We propose to use typically $k=2$ to 4, except when 
testing for independence. In the latter case we do not have to worry about
systematic errors, and statistical errors are minimized by taking $k$ to
be very large (up to $k\approx N/2$, say).

The above shows that $I^{(1)}(X,Y)$ and $I^{(2)}(X,Y)$ behave very 
similar. Also CPU times needed to estimate them are nearly the same.
In the following, we shall only show data for one of them, understanding 
that everything holds also for the other, unless the opposite is 
said explicitly.

\begin{figure}
  \begin{center}
    \psfig{file=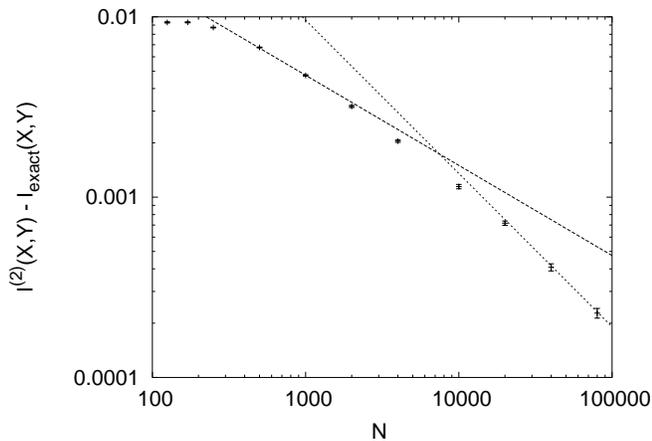,width=6cm,angle=270}
 \caption{Systematic error $I^{(2)}(X,Y)-I_{\rm exact}(X,Y)$ for $k=3$ plotted
    against $N$ on log-log scale, for $r=0.9$. The dashed lines are $\propto N^{-0.5}$ 
    and $\propto N^{-0.85}$.}
 \label{converge}
\end{center}
\end{figure}

For $N\to\infty$, the systematic errors tend to zero, as they should. From 
Figs.~2 and 4 one might conjecture that $I^{(1,2)}(X,Y) - I_{\rm exact}(X,Y)
\sim N^{-1/2}$, but this is not true. Plotting this difference on a double
logarithmic scale (Fig.~\ref{converge}), we see a scaling $\sim N^{-1/2}$ 
for $N\approx 10^3$, but faster convergence for larger $N$.
It can be fitted by a scaling $\sim 1/N^{0.85}$ for the largest 
values of $N$ reached by our simulations, but the true asymptotic behaviour 
is presumably just $\sim 1/N$.

\begin{figure}
  \begin{center}
    \psfig{file=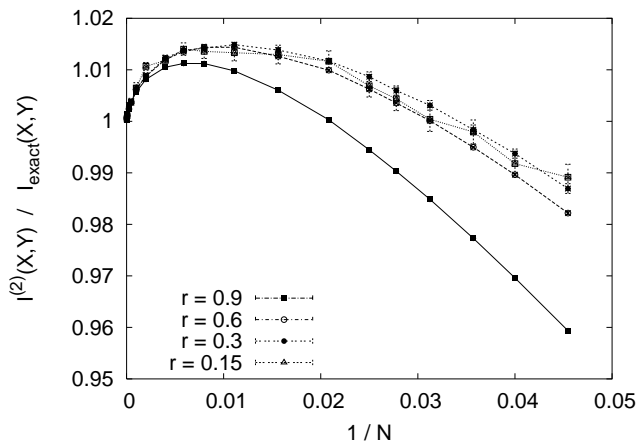,width=6cm,angle=270}
 \caption{Ratios $I^{(2)}(X,Y)/I_{\rm exact}(X,Y)$ for $k=1$ plotted
    against $1/N$, for 4 different values of $r$.}
 \label{conject}
\end{center}
\end{figure}

As said in the introduction, the most surprising feature of our estimators
is that they seem to be exact for independent random variables $X$ and $Y$.
In Fig.~\ref{conject} we show how the {\it relative} systematic errors
behave for Gaussians when $r\to 0$. More precisely, we show $I^{(1,2)}(X,Y)/
I^{(1,2)}_{\rm exact}(X,Y)$ for $k=1$, plotted against $N$ for four different 
values of $r$. Obviously these data converge, when $r\to 0$, to a finite function
of $N$. We have observed the same also for other distributions, which leads 
to a conjecture stronger than the conjecture made in the introduction:
Assume that we have a one-parameter family of 2-d distributions with 
densities $\mu(x,y;r)$, with $r$ being a real-valued parameter. Assume also
that $\mu$ factorizes for $r=r_0$, and that it depends smoothly on $r$
in the vicinity of $r_0$, with $\partial \mu(x,y;r)/\partial r$ finite.
Then we propose that for many distributions (although not for all!)
\be
   I^{(1,2)}(X,Y)/I_{\rm exact}(X,Y) \to F(k,N)
   \label{converge-r}
\ee
for $r\to r_0$, with some function $F(k,N)$ which is close to 1 for all $k$
and all $N\gg 1$, and which converges to 1 for $N\to \infty$. We have not 
found a general criterion for which families of distributions we should expect 
Eq.(\ref{converge-r}).

The most precise and efficient previous algorithm for estimating MI is the one 
of Darbellay \& Vajda \cite{darbellay-vajda}. As far as speed is concerned,
it seems to be faster than the present one, which might however 
be due to a more efficient implementation. In any case, also with the 
present algorithm we were able to obtain extremely high statistics on 
work stations within reasonable CPU times. To compare our statistical 
and systematic errors with those of \cite{darbellay-vajda}, we have used 
the code {\sf basic.exe} from {\sf http://siprint.utia.cas.cz/timeseries/}.
We used the parameter settings recommended in its description. 

\begin{figure}
  \begin{center}
    \psfig{file=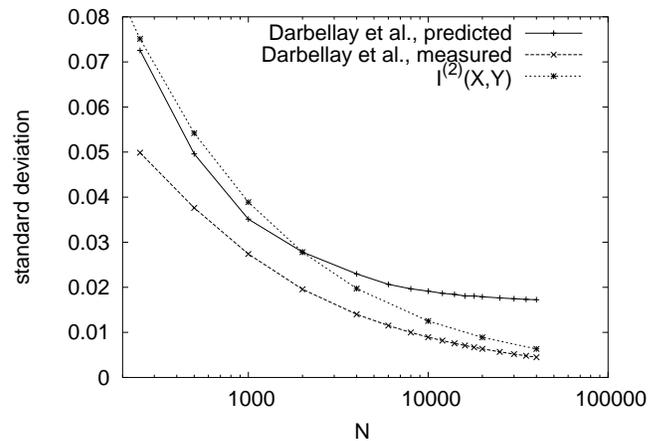,width=6cm,angle=270}
 \caption{Statistical errors (1 standard deviation) for Gaussian 
    deviates with $r=0.9$, plotted against $N$. Results from $I^{(2)}(X,Y))$ 
    for $k=1$ (full line) are compared to theoretically predicted (dashed) 
    and actually measured (dotted line) errors from \cite{darbellay-vajda}.}
 \label{stdev-darb}
\end{center}
\end{figure}

This code provides an estimate of the statistical error, even if only 
one data set is provided. When running it with many (typically $\approx 10^4$)
data sets, we found that these error bars are always underestimated, sometimes
by rather large margins. This seems to be due to occasional outliers which 
point presumably to some numerical instability. Unfortunately, having no
source code we could not pin down the troubles. In Fig.~\ref{stdev-darb} 
we compare the statistical errors provided by the code of \cite{darbellay-vajda},
the errors obtained from the variance of the output of this code, and 
the error obtained from $I^{(2)}(X,Y)$ with $k=3$. We see that the latter 
is larger than the theoretical error from \cite{darbellay-vajda}, but 
smaller than the actual error. For Gaussians with smaller correlation 
coefficients the statistical errors of \cite{darbellay-vajda} decrease
strongly with $r$, because the partitionings are followed to less 
and less depth. But, as we shall see, this comes with a risk for 
systematic errors.

\begin{figure}
  \begin{center}
    \psfig{file=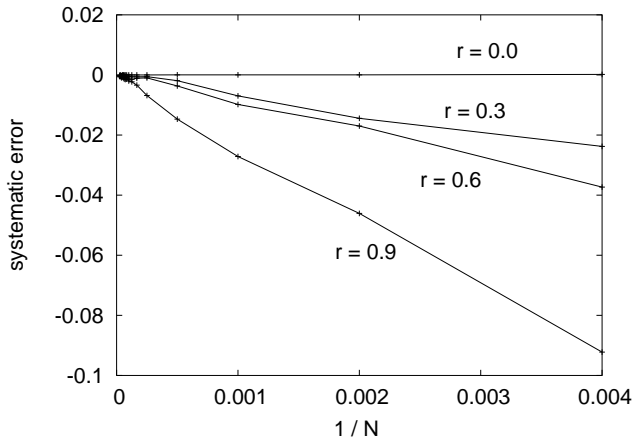,width=6cm,angle=270}
 \caption{Systematic errors for Gaussian 
    deviates with $r=0.0, 0.3, 0.6$, and $0.9$, plotted against $1/N$,
    obtained with the algorithm of \cite{darbellay-vajda}. These should 
    be compared to the systematic errors obtained with the present algorithm
    shown in Fig.~2.}
 \label{systdev-darb}
\end{center}
\end{figure}

Systematic errors of \cite{darbellay-vajda} for Gaussians with various 
values of $r$ are shown in Fig.~\ref{systdev-darb}. 
Comparing with Fig.~2 we see that they are, for $r\neq 0$, 
about an order of magnitude larger than ours, except for very large $N$
where they seem to decrease as $1/N$. Systematic errors of 
\cite{darbellay-vajda} are also very small when $r=0$, but this seems to 
result from fine tuning the parameter $\delta_s$ which governs the pruning 
of the partitioning tree in \cite{darbellay-vajda}. Bad choices of $\delta_s$
lead to wrong MI estimates, and optimal choices should depend on the 
problem to be analyzed. No such fine tuning is needed with our method.

As examples of non-Gaussian distributions we studied 
\begin{itemize}
\item The gamma-exponential distribution \cite{darb-vajd98};
\item The ordered Weinman exponential distribution \cite{darb-vajd98};
\item The ``circle distribution" of Ref.\cite{darb99}.
\end{itemize}
For all these, both exact formulas for the MI and detailed simulations
using the Darbellay-Vajda algorithm exist.
In addition we tested that $I^{(1)}$ and $I^{(2)}$ vanish, within 
statistical errors, for independent uniform distributions, for exponential 
distributions, and when $X$ was Gaussian and $Y$ was either uniform or 
exponentially distributed. Notice that `uniform' means uniform within 
a finite interval and zero outside, so that the Kozachenko-Leonenko 
estimate is not exact for this case either.

In all cases with independent $X$ and $Y$ we found that $I^{(1,2)}(X,Y)=0$
within the statistical errors (which typically were $\approx 10^{-3}$ to 
$10^{-4}$). We do not show these data.

\begin{figure}
  \begin{center}
    \psfig{file=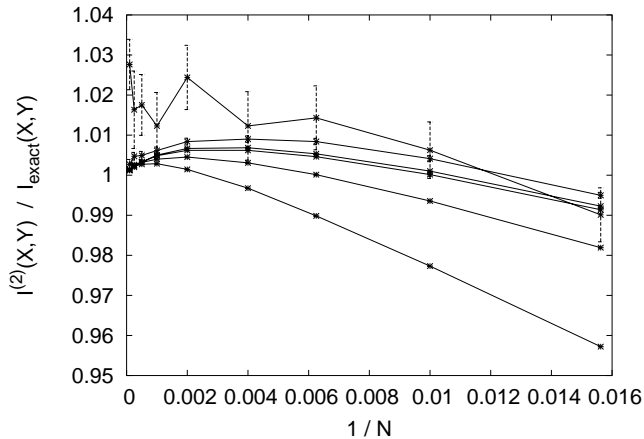,width=6cm,angle=270}
 \caption{Ratios $I(X,Y)_{\rm estim}/I_{\rm exact}(X,Y)$ for the 
    gamma-exponential distribution, plotted against $1/N$. 
    These data were obtained with $I^{(2)}$ using $k=1$, after 
    transforming $x_i$ and $y_i$ to their logarithms. The five
    curves correspond to $\theta = 0.1, 0.3, 1.0, 2.0, 10.0$, and 100.0 
    (from bottom to top).}
 \label{gamma-exp}
\end{center}
\end{figure}

\begin{figure}
  \begin{center}
    \psfig{file=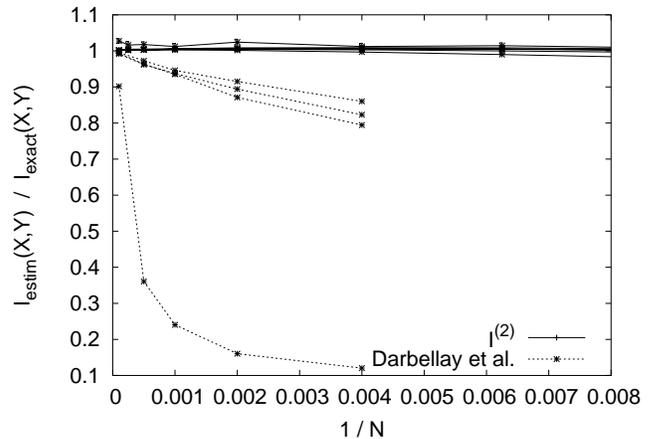,width=6cm,angle=270}
 \caption{Ratios $I(X,Y)_{\rm estim}/I_{\rm exact}(X,Y)$ for the gamma-exponential
    distribution, plotted against $1/N$. Full lines are from estimator 
    $I^{(2)}$, dashed lines are from \cite{darb-vajd98}. Our data were 
    obtained with $k=1$ after a transformation to logarithms. The four
    curves correspond to $\theta = 0.1, 0.3, 2.0$, and 100.0 (from bottom
    to top for our data, from top to bottom for the data of \cite{darb-vajd98}).}
 \label{gamma-exp-darb}
\end{center}
\end{figure}

The gamma-exponential distribution depends on a parameter $\theta$
(after a suitable re-scaling of $x$ and $y$) 
and is defined \cite{darb-vajd98} as
\be
   \mu(x,y;\theta) = {1\over \Gamma(\theta)} x^\theta e^{-x-xy}
\ee
for $x>0$ and $y>0$, and $\mu(x,y;\theta)=0$ otherwise. The MI is
\cite{darb-vajd98} $I(X,Y)_{\rm exact} = \psi(\theta+1)-\log \theta$. 
For $\theta>1$ the distribution
becomes strongly peaked at $x=0$ and $y=0$. Therefore, as we already
said, our algorithms perform poorly for $\theta \gg 1$, if we use $x_i$ 
and $y_i$ themselves. But using $x'_i = \log x_i$ and $y'_i = \log y_i$
we obtain excellent results, as seen from Fig.~\ref{gamma-exp}. There
we plot again $I^{(2)}(X',Y')/I(X,Y)_{\rm exact}$ for $k=1$ against $1/N$, 
for five values of $\theta$. These data obviously support our conjecture
that $I^{(2)}(X',Y')/I(X,Y)_{\rm exact}$ tends towards a finite function
as independence is approached. To compare with \cite{darb-vajd98}, we 
show in Fig.~\ref{gamma-exp-darb} our data together with those of 
\cite{darb-vajd98},
for the same four values of $\theta$ studied also there,
namely $\theta=0.1, 0.3, 2.0$, and $100.0$. 
We see that MI was grossly underestimated in \cite{darb-vajd98},
in particular for large $\theta$ where $I(X,Y)$ is very small (for
$\theta\gg 1$, one has $I(X,Y)\approx 1/2\theta$). 

\begin{figure}
  \begin{center}
    \psfig{file=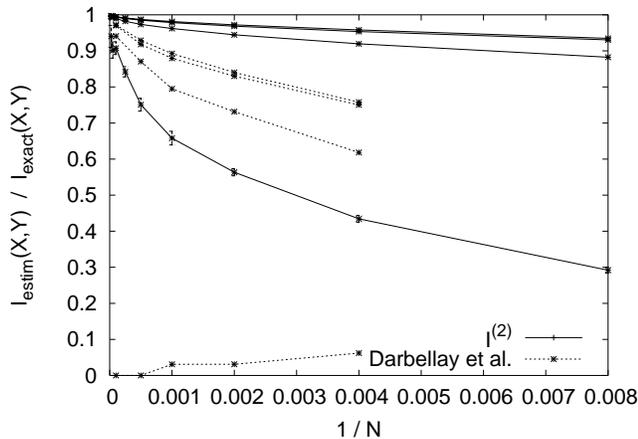,width=6cm,angle=270}
 \caption{Ratios $I(X,Y)_{\rm estim}/I_{\rm exact}(X,Y)$ for the ordered
    Weinman exponential
    distribution, plotted against $1/N$. Full lines are from estimator 
    $I^{(2)}$, dashed lines are from \cite{darb-vajd98}. Our data were 
    obtained with $k=1$ after a transformation to logarithms. The four
    curves correspond to $\theta = 0.1, 0.3, 1.0$, and 100.0 (from top
    to bottom).}
 \label{weinman}
\end{center}
\end{figure}

The ordered Weinman exponential distribution depends on two continuous
parameters. Following \cite{darb-vajd98} we consider here only the case 
where one of these parameters (called $\theta_0$ in \cite{darb-vajd98})
is set equal to 1, in which case the density is
\be
   \mu(x,y;\theta) = {2\over \theta} e^{-2x-(y-x)/\theta}
\ee
for $x>0$ and $y>0$, and $\mu(x,y;\theta)=0$ otherwise. The MI is
\cite{darb-vajd98} 
\be
   I(X,Y)_{\rm exact} = \left\{ \begin{array}{l@{\quad}l}
        \log{2\theta\over 1-2\theta} + \psi({1\over 1-2\theta}) -\psi(1) &:\quad \theta < {1\over 2}\\
                 & \\
        -\psi(1) &:\quad \theta ={1\over 2}\\
                 & \\
        \log{2\theta-1\over \theta} + \psi({2\theta\over 2\theta-1}) -\psi(1) &:\quad \theta > {1\over 2}
        \end{array} \right.
\ee
Mutual information estimates using $I^{(2)}(X,Y)$ with $k=1$ are shown 
in Fig.~\ref{weinman}. Again we transformed $(x_i,y_i) \to (\log x_i, \log y_i)$
since this improved the accuracy, albeit not as much as 
for the gamma-exponential distribution. More precisely, we plot 
$I^{(2)}(X,Y)/I(X,Y)_{\rm exact}$ against $1/N$ for the same four values
of $\theta$ studied also in \cite{darb-vajd98}, and we plot also the 
estimates obtained in \cite{darb-vajd98}. We see that MI 
was severely underestimated in \cite{darb-vajd98}, in particular
for large $\theta$ where the MI is small (for $\theta\to\infty$, one has
$I(X,Y)\approx (\psi'(1)-1)/2\theta = 0.32247/\theta$). Our estimates
are also too low, but much less so. It is clearly seen that
$I^{(2)}(X',Y')/I(X,Y)_{\rm exact}$ decreases for $\theta\to\infty$ in 
contradiction to the above conjecture. This represents
the only case where the conjecture does not hold numerically. As we 
already said, we do not know which feature of the ordered Weinman exponential
distribution is responsible for this difference.

\begin{figure}
  \begin{center}
    \psfig{file=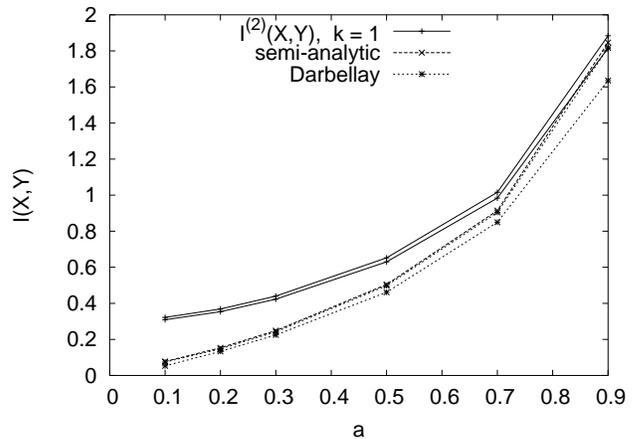,width=6cm,angle=270}
 \caption{Values of $I(X,Y)$ for the circle distribution of Ref.\cite{darb99},
    plotted against the inner radius parameter $a$. Full lines are 
    from estimator $I^{(2)}$ with $k=1$ and with $N=1000$ (bottom) and
    $N=10000$ (top). Dashed lines are estimates from Ref.\cite{darb99}
    with $N=1000$ and $N=10000$, and the semi-analytical result of 
    Ref.\cite{darb99} (also from bottom to top).}
 \label{circle}
\end{center}
\end{figure}

The `circle distribution' of Ref.\cite{darb99} is defined in terms of polar
coordinates $(r,\phi)$ as uniform in $\phi$, and with a triangular radial
profile: The radial distribution $p_R(r)$ vanishes for $r<a$ and $r>1$, 
is maximal at $r=(1+a)/2$, and is linear in the intervals $[a,(1+a)/2]$
and $[(1+a)/2,1]$. The variables $X$ and $Y$ are obtained as $X=r\cos\phi$
and $Y=r\sin\phi$. In Ref.\cite{darb99} it was shown that its MI can be
calculated analytically except for one integral which has to be done 
numerically. Tables for the exact and estimated values of MI in the range
$0.1 \leq a \leq 0.9$ are given in Ref.\cite{darb99}, from which it appears
that the Darbellay-Vajda algorithm is very precise for this case.
Unfortunately, our own estimates (using again $I^{(2)}$ with $k=1$) are in 
serious disagreement with them, see Fig.~\ref{circle}. Moreover, the values 
quoted in Ref.\cite{darb99} seem to converge to $I(X,Y)\to 0$ for $a\to 0$
which is impossible ($X$ and $Y$ are not independent even for $a=0$). 
We have no explanation for this. In any case, our estimates are rapidly 
convergent for $N\to\infty$.

\subsection{Higher Dimensions}

\begin{figure}
  \begin{center}
    \psfig{file=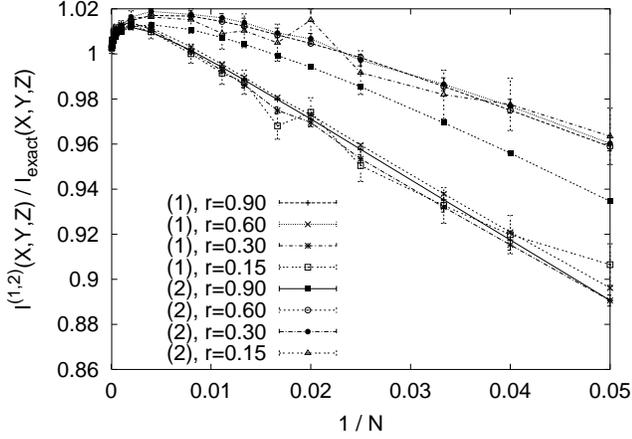,width=6cm,angle=270}
 \caption{Ratios $I^{(1,2)}(X,Y,Z)/I_{\rm exact}(X,Y,Z)$ for $k=1$ plotted
    against $1/N$, for 4 different values of $r$. All Gaussians have
    unit variance and all non-diagonal elements in the correlation matrix
    $\sigma_{i,k}, i\not=k$  (correlation coefficients) take the value
    $r$.  }
 \label{conject3}
\end{center}
\end{figure}

In higher dimensions we shall only discuss applications of our
estimators  to {\it m} correlated Gaussians, because as in the case of
two dimensions this is easily compared to analytic results
(Eq.(\ref{gauss-analytic})) and to previous  numerical results \cite{darbHD}. 
As already mentioned in the introduction and as shown above for 2-d 
distributions (Fig.~\ref{conject}) our estimates seem to be exact for
independent random variables. We choose the same one-parameter family of 3-d
Gaussian distributions with all the correlation coefficients equal to
$r$ as in \cite{darbHD}. In Fig.~\ref{conject3} we show  the behavior of the {\it
relative} systematic errors of both proposed estimators. 
One can easily  see that the data converge for $r\to0$, i.e. when all three
Gaussians become independent.  This supports the conjecture made in the 
previous subsection.
In addition, in Fig.~\ref{conject3} one can see the difference between the 
estimators $I^{(1)}$ and $I^{(2)}$. For intermediate numbers of the points,
$N\sim 100-200$, the ``cubic" estimator has lower systematic error. Apart from 
that, $I^{(2)}$ evaluated for $N$ is roughly equal to $I^{(1)}$ evaluated
for $2N$, reflecting the fact that $I^{(2)}$ effectively uses smaller length
scales as discussed already for $d=2$.

\begin{figure}
  \begin{center}
    \psfig{file=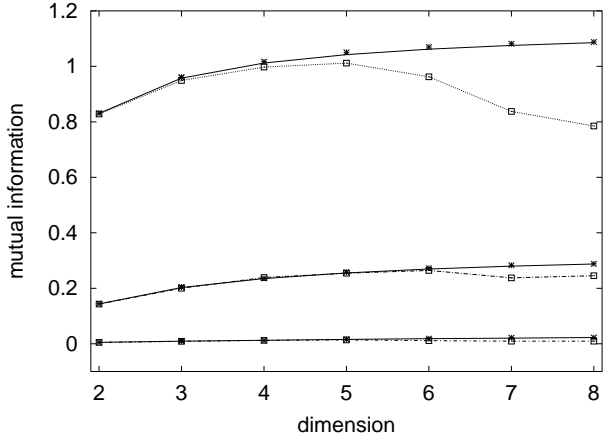,width=6cm,angle=270}
 \caption{Averages of $I^{(1,2)}(X_1,(X_2...X_m))$ for $k=1$ plotted
    against $d$, for 3 different values of $r= 0.1, 0.5, 0.9$. The
    sample size is 50000, averaging is done over 100 realizations
    (same parameters as in \cite{darbHD}, Fig. 1).  Full lines indicate 
    theoretical values, pluses $(+)$ are for $I^{(1)}$, crosses $(\times)$ for 
    $I^{(2)}$. Squares and dotted lines are read off from 
    Fig. 1 of Ref.\cite{darbHD}.}
 \label{mihighdim}
\end{center}
\end{figure}

To compare our results in high dimension with the ones presented in
\cite{darbHD} we shall calculate not the high dimensional redundancies
$I(X_1,X_2,...,X_m)$ but the MI  $I((X_1,X_2,...,X_{m-1}),X_m)$ between
two variables, namely an $m-1$ dimensional vector and a scalar. For
estimation of this MI we can use the formulas as for the 2-d case (Eq.(\ref{i1}) 
and Eq.(\ref{i2}), respectively) where $n_x$ would be defined as the number
of points in the ($m-1$)-dimensional stripe of (hyper-)cubic cross section.
Using directly 
Eq.(\ref{a2}) would increase the errors in  estimation (see the appendix for 
the relation between $I(X_1,X_2,...,X_m)$ and $I((X_1,X_2,...,X_{m-1}),X_m)$).

In Fig.~\ref{mihighdim} we show the average values of $I^{(1,2)}$. 
They are in very good agreement with the theoretical ones for all three 
values of the correlation coefficient $r$ and all dimensions tested here
(in contrast, in \cite{darbHD} the estimators of MI significantly deviate 
from the theoretical values for dimensions $\geq 6$).
It is impossible to distinguish (on this scale) between
estimates $I^{(1)}$ and $I^{(2)}$. 

\begin{figure}
  \begin{center}
    \psfig{file=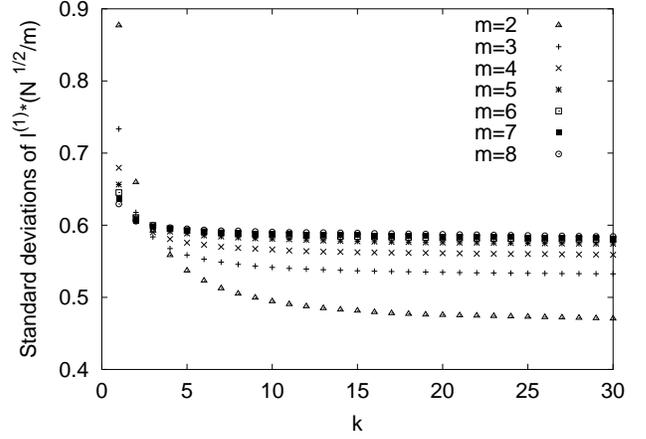,width=6cm,angle=270}
    \caption{Standard deviations of the estimate $I^{(1)}$ 
      for Gaussian deviates with unit variance and covariance $r=0.9$,
      multiplied by $\frac{\sqrt{N}}{m}$ and plotted  against
      $k$. Each curve corresponds to a fixed value of dimension
      $m$. Number of samples is $N=10000$.}
    \label{stdhd}
\end{center}
\end{figure}

In Fig.~\ref{stdhd}, statistical errors of our estimate are presented as a
function of the number of neighbours $k$.  More precisely, we plotted the 
standard deviation of $I^{(1)}$ multiplied by $\frac{\sqrt{N}}{m}$ against
$k$ for the case where all correlation coefficients are $r=0.9$. Each curve 
corresponds to a different  dimension $m$. The data scale roughly as $\sim
\frac{m}{\sqrt{N}}$ for large dimension. Moreover, these statistical errors 
seem to converge to finite values for $k\to\infty$. This convergence 
becomes faster for increasing dimensions. The same behavior is observed for $I^{(2)}$.

\section{Applications: Gene Expression Data and Independent Component Analysis}

\subsection{Gene Expression}

In the first application to real world data, we compare our MI estimators to
kernel density estimators (KDE) made in \cite{steuer02}. The data there are gene 
expression data from \cite{hughes00}. They consist of sets of $\approx 300$ vectors
in a high dimensional space, each point corresponding to one genome and each 
dimension corresponding to one open reading frame (ORF). Mutual informations 
between data corresponding to pairs of ORFs (i.e., for 2-d projections of the 
set of data vectors) are estimate in \cite{steuer02} to improve eventually the 
hierarchical clustering made in \cite{hughes00}. It was found that KDE performed much
better than estimators based on binning, but that 
the estimated MIs were so strongly correlated to linear correlation coefficients
that they hardly carried more useful information.

Here we re-investigate just the MI estimates of the four ORF pairs ``A" to 
``D" shown in Figs. 3, 5, and 7 of \cite{steuer02}. The claim that KDE was 
superior to binning was based on a surrogate analysis. For surrogates consisting 
of completely independent pairs, KDE was able to show that all four pairs 
were significantly dependent, while binning based estimators could disprove
the null hypothesis of independence only for two pairs. In addition, KDE had 
both smaller statistical and systematic errors. Both KDE and binning estimators
were applied to rank-ordered data \cite{steuer02}.

In KDE, the densities are approximated by sums of $N$ Gaussians with fixed 
prescribed width $h$ centered at the data points. In the limit $h\to 0$ the 
estimated MI diverges, while it goes to zero for $h\to\infty$. Our main 
criticism of \cite{steuer02} is that the authors used a very large value
of $h$ (roughly 1/2 to 1/3 of the total width of the distribution). This is 
recommended in the literature \cite{silver86}, since both statistical and 
systematic errors would become too large for smaller values of $h$. But with
such a large value of $h$ one is insensitive to finer details of the distributions,
and it should not surprise that hardly anything beyond linear correlations is 
found by the analysis.

With our present estimators $I^{(1)}$ and $I^{(2)}$ we found indeed considerably
larger statistical errors, when using small values of $k$ ($k<10$, say). But 
when using $k\approx 50$ (corresponding to $\sqrt{k/N}\approx 0.4$, similar to
the ratio $h/\sigma$ used in \cite{steuer02}) the statistical errors were 
comparable to those in \cite{steuer02}. Systematic errors could be 
estimated by using the exact inequality Eq.(\ref{bound}) given in the appendix
(when applying this, one has of course to remember that the estimate of the 
correlation coefficient contains errors which lead to systematic overestimation
of the r.h.s. of Eq.(\ref{bound}) \cite{darbellay-vajda}). For instance, for 
pair ``B" one finds $I(X,Y)>1.1$ from Eq.(\ref{bound}). While this is satisfied
for $k<5$ within the expected uncertainty, it is violated both by the estimate
of \cite{steuer02} ($I(X,Y)\approx 0.9$) and by our estimate for $k=50$ 
($I(X,Y)\approx 0.7$). With our method and with $k\approx 50$, we could also
show that none of the four pairs is independent, with roughly the same 
significance as in \cite{steuer02}.

Thus the main advantage of our method is that it does not deteriorate as quickly
as KDE does for high resolution. In addition, it seems to be faster, although
the precise CPU time depends on the accuracy of the integration needed in KDE.
In \cite{steuer02} also a simplified algorithm is given (Eq.(33) of 
\cite{steuer02}) where the integral is replaced by a sum. Although it is 
supposed to be faster that the algorithm involving numerical integration
(on which were based the above estimates), it is much slower that our present 
estimators (it is $O(N^2)$ and involves the evaluation of $3N^2$ exponential 
functions). This simplified algorithm (which is indeed just a generalized
correlation sum with the Heaviside step function replaced by Gaussians) gives
also rather big systematic errors, e.g. $I(X,Y)= 0.66$ for pair ``B".

\subsection{ICA}

Independent component analysis (ICA) is a statistical method for transforming
an observed multi-component data set (e.g. a multivariate time series comprising 
$n$ measurement channels) 
${\bf x}(t)=(x_1(t),x_2(t),...,x_n(t))$ into components that are statistically
as independent from each other as possible \cite{hyvar2001}. In the simplest 
case, ${\bf x}(t)$ could be a linear superposition of $n$ independent sources
${\bf s}(t)=(s_1(t),s_2(t),...,s_n(t))$,
\be
  {\bf x}(t) =  {\bf A}\;{\bf s}(t)\; ,
  \label{source-x}
\ee
where ${\bf A}$ is a non-singular $n\times n$ `mixing matrix'.
In that case, we know that a decomposition into independent components is 
possible, since the inverse transformation 
\be
   {\bf s}(t) = {\bf Wx}(t) \;\;\text{ with } \;\; {\bf W}={\bf A}^{-1}
  \label{x-source}
\ee
does exactly this. If Eq.(\ref{source-x}) does not hold, then no decomposition
into strictly independent components is possible by a linear transformation 
like Eq.(\ref{x-source}), but one can still search for least dependent 
components. In a slight misuse of notation, this is still called ICA.

But even if Eq.(\ref{source-x}) does hold, the problem of blind source 
separation (BSS), i.e. finding the matrix ${\bf W}$ without explicitly knowing 
${\bf A}$, is not trivial. Basically, it requires that ${\bf x}$ is such that
all superpositions ${\bf s}' = {\bf W}'{\bf x}$ with ${\bf W}'\ne {\bf W}$
are not independent. Since linear combinations of Gaussian variables are also 
Gaussian, BSS is possible only if the sources are not Gaussian. Otherwise, any 
rotation (orthogonal transformation) ${\bf s}' = {\bf Rs}$ would again lead 
to independent components, and the original sources ${\bf s}$ could not be 
uniquely recovered.

This leads to basic performance tests for any ICA problem:\\
1) How independent are the found ``independent" components?\\
2) How unique are these components?\\
3) How robust are the estimated {\it dependencies} against noise, against statistical 
   fluctuations, and against outliers?\\
4) How robust are the estimated {\it components}?

Different ICA algorithms can then be ranked by how well they perform, 
i.e. whether they find indeed the most independent components, 
whether they declare them as unique if and only if they indeed are, 
and how robust are the results.
While questions 2 and 4 have often been discussed in the ICA literature
(for a particularly interesting recent study, see \cite{Meinecke02}), the first
(and most basic, in our opinion) test has not attracted much interest. This
might seem strange since MI is an obvious candidate for measuring independence,
and the importance of MI for ICA was noticed from the very beginning. We
believe that the reason was the lack of good MI estimators. We propose to 
use our MI estimators not only for testing the actual independence of the 
components found by standard ICA algorithms, but also to use them for testing
for uniqueness and robustness. We will also show how our estimators can be 
used for improving the decomposition obtained from a standard ICA algorithm, 
i.e. for finding components which are more independent. Algorithms which 
use our estimators for ICA from scratch will be discussed elsewhere.

It is useful to decompose the matrix {\bf W} into two factors, ${\bf W} = {\bf R}{\bf V}$,
where {\bf V} is a prewhitening that transforms the covariance matrix into ${\bf C}' = 
{\bf V}{\bf C}{\bf V}^T = {\bf 1}$, and {\bf R} is a pure rotation. Finding and applying 
{\bf V} is just a principal component analysis (PCA) together with a rescaling, 
so the ICA problem proper reduces to finding
a suitable rotation after having the data prewhitened. In the following we always
assume that the prewhitening (PCA) step has already been done.

Any rotation can be represented as a product of rotations which act only in some 
$2\times 2$ subspace, ${\bf R} = \prod_{i,j} {\bf R}_{ij}(\phi)$, where 
\be
   {\bf R}_{ij}(\phi)(x_1,\ldots x_i\ldots x_j\ldots x_n) = (x_1,\ldots x_i'\ldots x_j'\ldots x_n)
   \label{R}
\ee
with 
\be
   x_i' = \cos\phi\; x_i + \sin\phi\; x_j,\quad x_j' = -\sin\phi\; x_i + \cos\phi\; x_j\;.
\ee
For such a rotation one has (see appendix)
\be
   I({\bf R}_{ij}(\phi){\bf X}) - I({\bf X}) = I(X_i',X_j') - I(X_i,X_j)\;,
   \label{change-I}
\ee
i.e. the change of $I(X_1\ldots X_n)$ under any rotation can be computed by adding up 
changes of two-variable MIs. This is an important numerical simplification. It would 
not hold if MI is replaced by some other similarity measure, and it indeed is not 
strictly true for our estimates $I^{(1)}$ and $I^{(2)}$. But we found the violations
to be so small that Eq.(\ref{change-I}) can still be used when minimizing MI.

Let us illustrate the application of our MI estimates to 
a fetal ECG recorded from the abdomen and thorax of a pregnant woman
(8 electrodes, 500 Hz, 5s). We chose this data set because it was analyzed by several
ICA methods \cite{Cardoso98,Meinecke02} and is available in the web \cite{ECGdata}.
In particular, we will use both $I^{(1)}$ and $I^{(2)}$ to check and improve the output 
of the JADE algorithm \cite{Cardoso93} (which is a standard ICA algorithm and was more 
successful with these data than TDSEP \cite{Ziehe98}, see \cite{Meinecke02}).

\begin{figure}
  \begin{center}
    \psfig{file=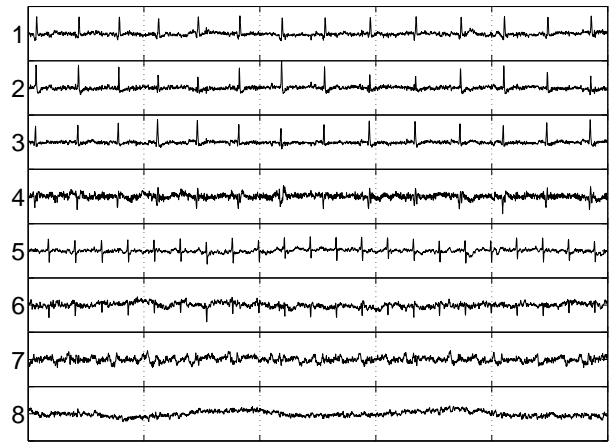,width=8.0cm,angle=0}
  \caption{Estimated independent components using JADE.}
 \label{ECGJADE}
 \end{center}
\end{figure}

\begin{figure}
  \begin{center}
    \psfig{file=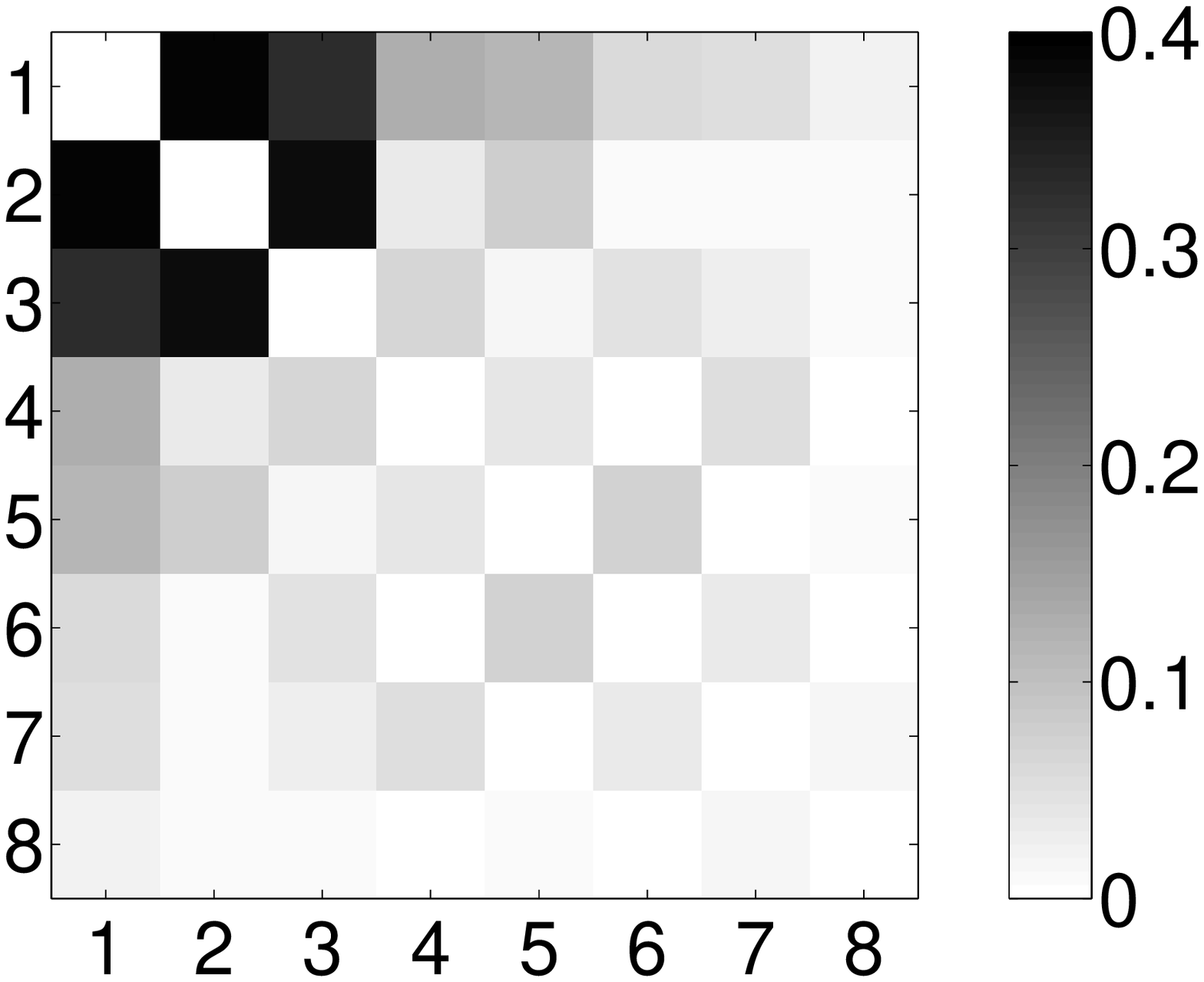,width=4.1cm,angle=0}
    \psfig{file=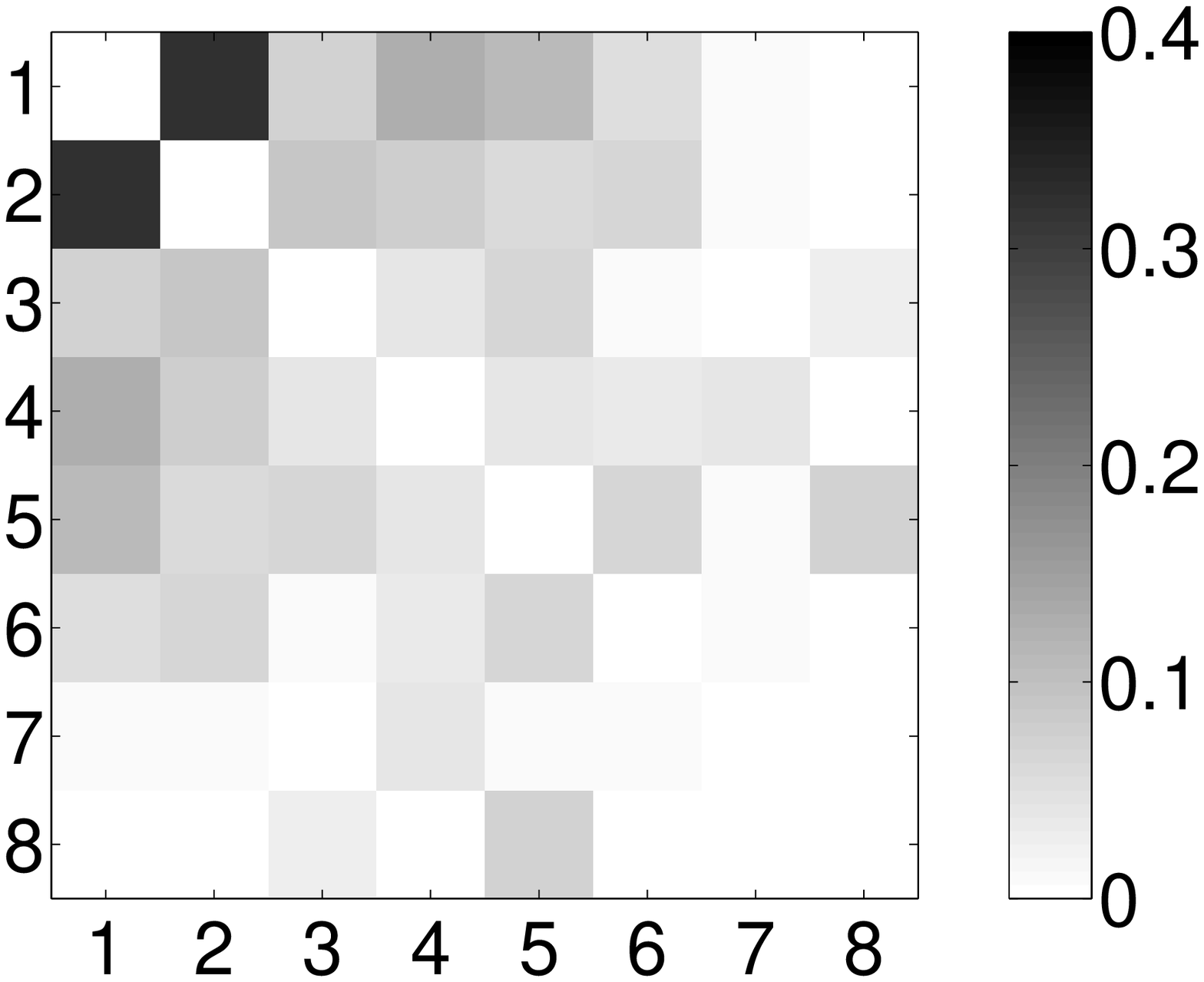,width=4.1cm,angle=0}
  \caption{Left panel: Pairwise MIs between all ICA-components obtained by JADE, 
     estimated with $I^{(1)}, k= 1$. The diagonal is set to
     zero.\\
     Right panel: Pairwise MIs between the optimized channels shown in
     Fig. \ref{ECGMIICA}.}
 \label{ECGmimatrix}
 \end{center}
 \end{figure}

The output of JADE for these data, i.e. the supposedly least dependent components,
are shown in Fig.~\ref{ECGJADE}. Obviously channels 1-3 are dominated by the 
heartbeat of the mother, and channel 5 by that of the child. Channels 4 and 6 still 
contain large heartbeat components (of mother and child, respectively), but look 
much more noisy. Channels 7-8 seem to be dominated by noise, but with rather different
spectral composition. The pairwise MIs of these channels are shown in Fig.~\ref{ECGmimatrix}
(left panel)\cite{footnote}. One sees that most MIs are indeed small, but the first 3 components
are still highly interdependent. This could be a failure of JADE, or it could mean
that the basic model does not apply to these components. To decide between these
possibilities, we minimized $I(X_1\ldots X_8)$ by means of Eqs.(\ref{R}) - (\ref{change-I}). 
For each pair $(i,j)$ with $i,j = 1\ldots 8$ we found the angle which minimized 
$I(X_i',X_j') - I(X_i,X_j)$, and repeated this altogether $\approx 10$ times. We 
did this both for $I^{(1)}$ and $I^{(2)}$, with $k = 1$. We checked that 
$I(X_1\ldots X_8)$, calculated directly, indeed decreased (from $I^{(1)}_{JADE} = 
1.782$ to $I^{(1)}_{\rm min} = 1.160$, and from $I^{(2)}_{JADE} = 2.264$ to 
$I^{(2)}_{\rm min} = 1.620$). 

\begin{figure}
  \begin{center}
    \psfig{file=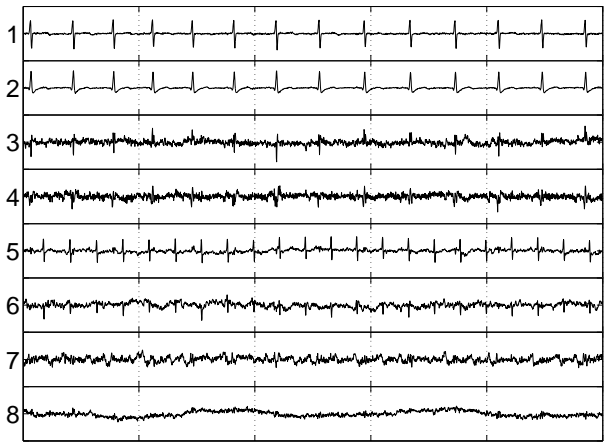,width=8.0cm,angle=0}
  \caption{Estimated independent components after minimizing $I^{1}$.}
 \label{ECGMIICA}
 \end{center}
\end{figure}

The resulting components are shown in Fig.~\ref{ECGMIICA}. The first 2 components 
look now much cleaner, all the noise from the first 3 channels seems now concentrated 
in channel 3. But otherwise things have not changed very much. The pairwise MI
after minimization are shown in Fig.~\ref{ECGmimatrix} (right panel). As suggested
by Fig.~\ref{ECGMIICA}, channel 3 is now much less dependent on channels 1 and 2.
But the latter are still very strongly interdependent, and a linear superposition of 
independent sources as in Eq.(\ref{source-x}) can be ruled out. This was indeed to be 
expected: In any oscillating system there must be at least 2 mutually dependent 
components involved, and generically one expects both to be coupled to the output
signal.

\begin{figure}
  \begin{center}
    \psfig{file=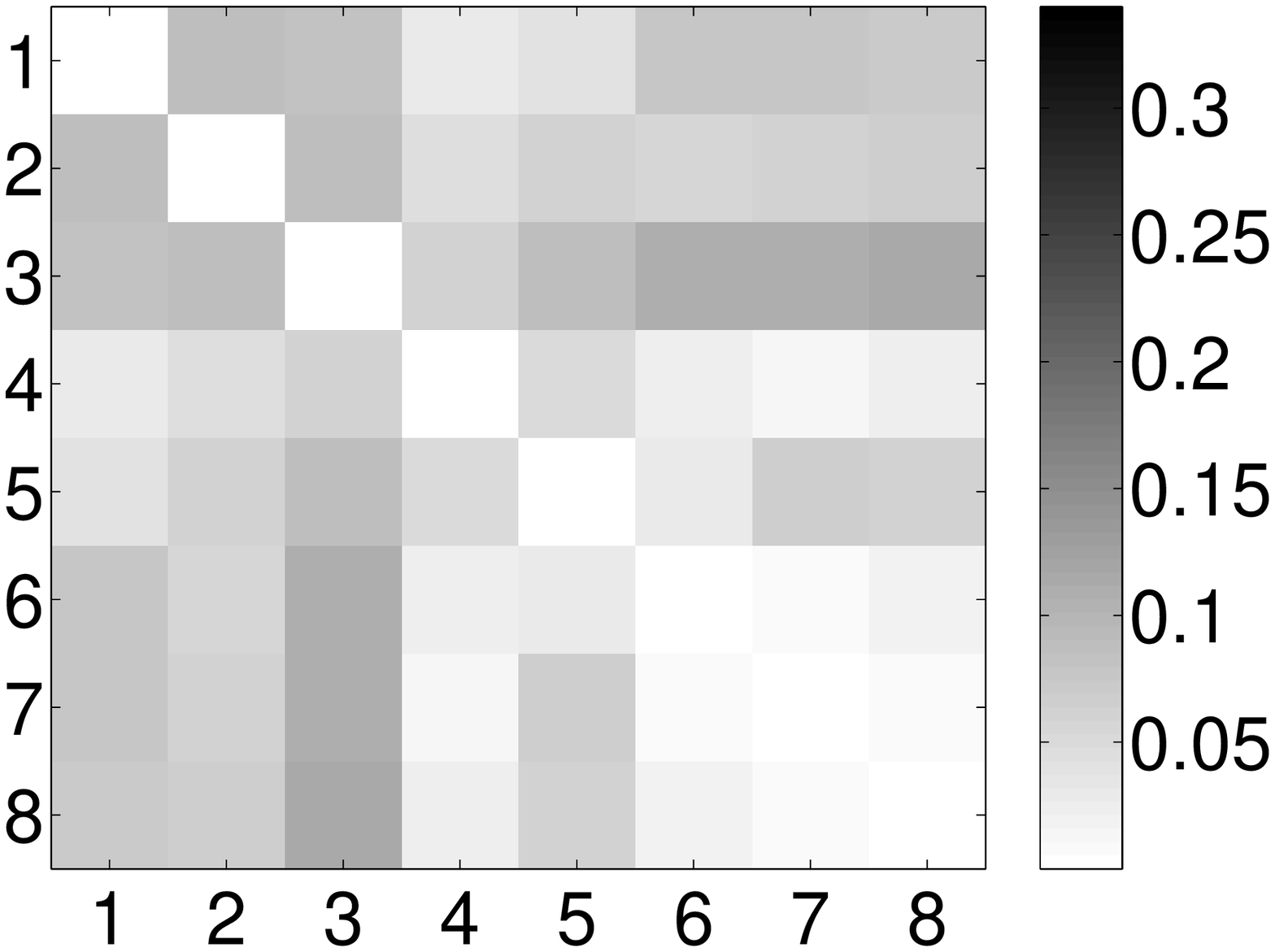,width=4.1cm,angle=0}
    \psfig{file=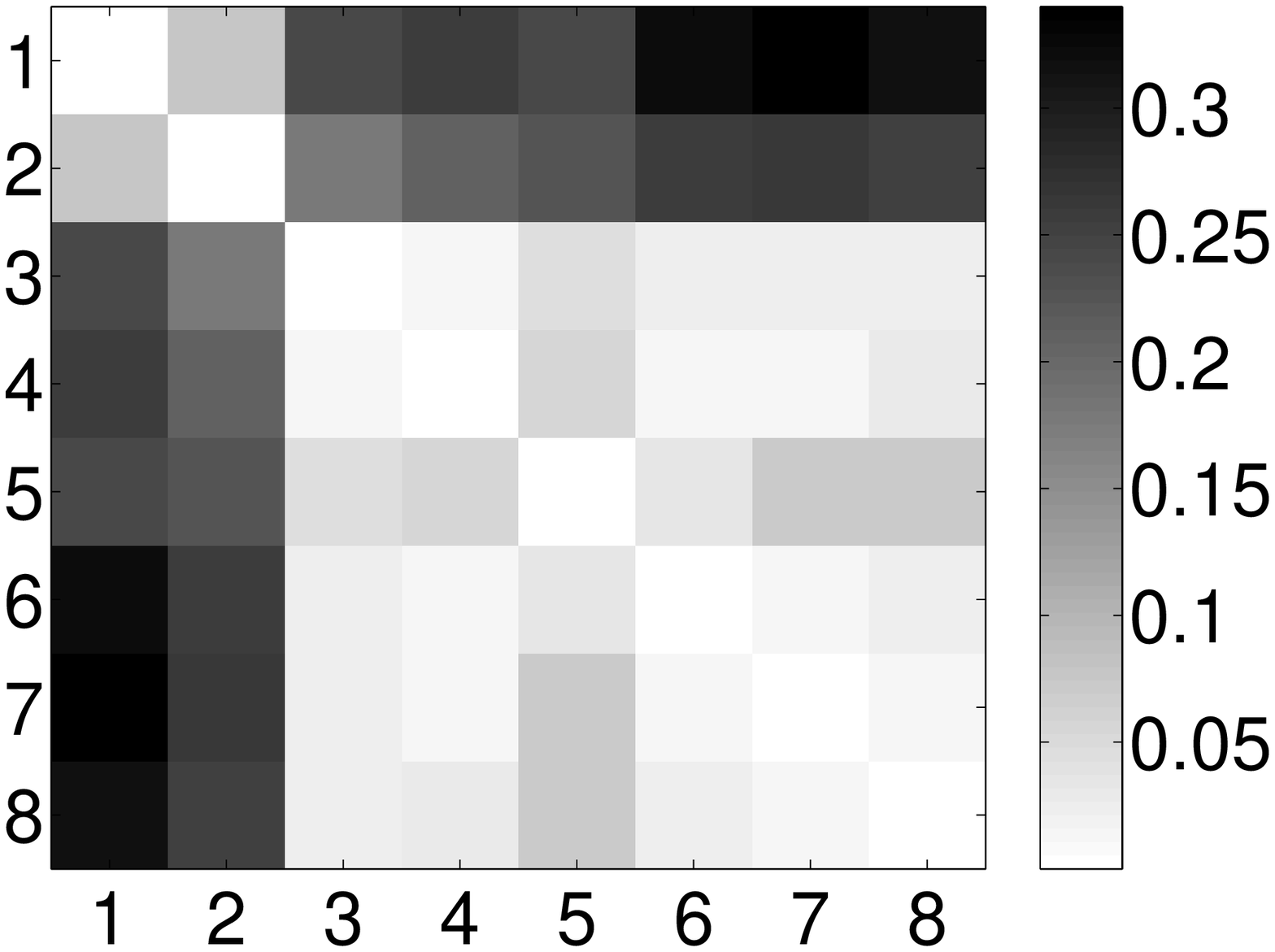,width=4.1cm,angle=0}
  \caption{Square roots of variances, $\sqrt{\sigma_{ij}}$, of 
     $I^{(1)}((X_i, X_j))$ (with $k=1$) from JADE output
     (left panel) and after minimization of MI (right panel). Again, elements
     on the diagonal have been put to zero.}
 \label{ECGmimatrix2}
 \end{center}
\end{figure}

To test for the uniqueness of the decomposition, we computed the variances
\be
   \sigma_{ij} = {1\over 2\pi}\int_0^{2\pi} d\phi \left[ I({\bf R}(\phi)(X_i,X_j))
    - \overline{I(X_i,X_j)}\right]^2\;,
\ee
where
\be
   \overline{I(X_i,X_j)} = {1\over 2\pi}\int_0^{2\pi} d\phi I({\bf R}(\phi)(X_i,X_j))\;.
\ee
If $\sigma_{ij}$ is large, the minimum of the MI with respect to rotations is 
deep and the separation is unique and robust. If it is small, however, BSS can
not be achieved since the decomposition into independent components is not robust.
Results for the JADE output are shown in Fig.~\ref{ECGmimatrix2} (left panel), 
those for the optimized decomposition in the right panel of Fig.~\ref{ECGmimatrix2}.
The most obvious difference between them is that the first two channels have become
much more clearly distinct and separable from the rest, while channel 3 is less
separable from the rest (except from channel 5). This makes sense, since 
channels 3, 4, 7, and 8 now contain mostly Gaussian noise which is featureless
and thus rotation invariant after whitening. Most of the signals are now contained
in channels 5 (fetus) and in channels 1 and 2 (mother).

These results are in good agreement with those of \cite{Meinecke02}, but are obtained
with less numerical effort and can be interpreted more straightforwardly.

\section{Conclusion}

We have presented two closely related families of mutual entropy estimators. In general
they perform very similarly, as far as CPU times, statistical errors, and systematic
errors are concerned. Their biggest advantage seems to be in vastly reduced systematic
errors, when compared to previous estimators. This allows us to use them on very
small data sets (even less than 30 points gave good results). It also allows us to 
use them in independent component analyses to estimate absolute values of mutual 
dependencies. Traditionally, contrast functions have been used in ICA which allow
to minimize MI, but not to estimate its absolute value. We expect that our 
estimators will also become useful in other fields of time series and pattern
analysis. One large class of problems is interdependencies in physiological time 
series, such as breathing and heart beat, or in the output of different EEG
channels. The latter is particularly relevant for diseases characterized by 
abnormal synchronization, like epilepsy or Parkinson's disease. In the past, various
measures of interdependence have been used, including MI. But the latter was not 
employed extensively (see, however, \cite{pompe}), mainly because of the supposed 
difficulty in estimating it 
reliably. We hope that the present estimators might change this situation.

\vskip 4mm

\noindent Acknowledgment:\\
One of us (P.G.) wants to thank Georges Darbellay for extensive and very fruitful 
e-mail discussions. We also want to thank Ralph Andrzejak, Thomas Kreuz, and Walter
Nadler for numerous fruitful discussions, and for critically reading the manuscript.

\section*{Appendix}

We collect here some well known facts about MI, in particular for higher 
dimensions, and some immediate consequences. The first important property of $I(X,Y)$ 
is its independence with
respect to reparametizations. If $X'=F(X)$ and $Y'=G(Y)$ are homeomorphisms
(smooth and uniquely invertible maps), and $J_X = ||\partial X/\partial X'||$
and $J_Y = ||\partial Y/\partial Y'||$ are the Jacobi determinants, then 
\be
   \mu'(x',y') = J_X(x') J_Y(y') \mu(x,y)
\ee
and similarly for the marginal densities, which gives
\bea
   I(X',Y') & = &\int\!\!\!\int dx' dy' \mu'(x',y') \log{\mu'(x',y')
       \over \mu'_x(x')\mu'_y(y')}  \nonumber \\
   & = & \int\!\!\!\int dx dy \;\mu(x,y)\; \log{\mu(x,y)\over \mu_x(x)\mu_y(y)} \nonumber \\
   & = & I(X,Y)\;.
   \label{a1}
\eea

The next important property, checked also directly from the definitions,
is 
\be
   I(X,Y,Z) = I((X,Y),Z) + I(X,Y)\;.
   \label{a2}
\ee
This is analogous to the additivity axiom for Shannon entropies \cite{cover-thomas},
and says that MI can be decomposed into hierarchical levels. By iterating it,
one can decompose $I(X_1\ldots X_n)$ for any $n>2$ and for any partitioning
of the set $(X_1\ldots X_n)$ into the MI between elements within
one cluster and MI between clusters.

Let us now consider a homeomorphism $(X',Y') = F(X,Y)$.
By combining Eqs.(\ref{a1}) and (\ref{a2}) we obtain
\bea
   I(X',Y',Z)& =& I((X',Y'),Z) + I(X',Y') \nonumber \\
             &= &I((X,Y),Z) + I(X',Y')              \\
             &= &I(X,Y,Z) + [I(X',Y')-I(X,Y)]\;. \nonumber
\eea
Thus, changes of high dimensional redundancies under reparametrization of some
subspace can be obtained by calculating MIs in this subspace only. Although
this is a simple consequence of well known facts about MI, it seems to have 
not been noticed before. It is numerically extremely useful, and would not hold 
in general for other interdependence measures. Again it
generalizes to any dimension and to any number of random variables. 

It is well known that Gaussian distributions maximize the Shannon entropy for
given first and second moments. This implies that the Shannon 
entropy of any distribution is bounded from above by $(1/2) \log \det {\bf C}$
where ${\bf C}$ is the covariance matrix. For MI one can prove a similar result:
For any multivariate distribution with joint covariance matrix ${\bf C}$ and 
variances $\sigma_i = C_{ii}$ for the individual (scalar) random variables $X_i$, 
the redundancy is bounded from below,
\be
   I(X_1,\ldots X_m) \geq {1\over 2} \log{\det {\bf C}\over \sigma_1\ldots
   \sigma_m} \;.
   \label{bound}
\ee
The r.h.s. of this inequality is just the redundancy of the corresponding 
Gaussian, and to prove Eq.(\ref{bound}) it we must show that the distribution 
minimizing the MI is Gaussian.

In the following we sketch only the proof for the case of 2 variables $X$ and 
$Y$, the generalization to $m>2$ being straightforward. We also assume without
loss of generality that $X$ and $Y$ have zero mean.
To prove Eq.(\ref{bound}), we set up a minimization problem where the 
constraints (correct normalization and correct second moments; 
consistency relations $\mu_x(x)=\int dy\;\mu(x,y)$ and 
$\mu_y(y)=\int dx\;\mu(x,y)$) are taken into account by means of Lagrangian
multipliers. The ``Lagrangian equation" $\delta L / \delta \mu(x,y) = 0$ 
leads then to 
\be
   \mu(x,y) = {1\over Z}\;\mu_x(x) \; \mu_y(y)\; e^{-ax^2-by^2-cxy}
   \label{appq}
\ee
where $Z,a,b,$ and $c$ are constants fixed by the constraints. Since the minimal 
MI decreases when the variances $\sigma_x = C_{xx}$ and $\sigma_y = C_{yy}$
increase with $C_{xy}$ fixed, the constants $a$ and $b$ are non-negative.
Eq.(\ref{appq}) is obviously consistent with $\mu(x,y)$ being a Gaussian. 
To prove uniqueness, we integrate Eq.(\ref{appq}) over $y$ and put $x=-iz/c$, 
to obtain
\be
   Z\; e^{-az^2/c^2} = \int dy \; e^{izy} \;[\mu_y(y) e^{-by^2}]\;.
\ee
This shows that $e^{-by^2}\mu_y(y)$ is the Fourier transform of a Gaussian, and 
thus $\mu_y(y)$ is also Gaussian. The same holds of course true for $\mu_x(x)$, 
showing that the minimizing $\mu(x,y)$ must be Gaussian, QED.

Finally, we should mention some possibly confusing notations. First, 
MI is often also called transinformation or redundancy. Secondly, what we call 
higher order redundancies are called higher order MIs in the ICA literature.
We did not follow that usage in order to avoid confusion with cumulant-type higher 
order MIs \cite{matsuda}.


\begin{thebibliography}{30}
\bibitem{footnote} In Figs.~18 to ~20 we used $k=1$, since the time sequences were
   sufficiently long to give very small statistical errors. To find components as
   independent as possible, we should have used much larger $k$, since this would 
   reduce statistical errors at the cost of increased but anyhow very small 
   systematic errors. We checked that basically the same results were obtained 
   with $k$ up to 100.
\end{thebibliography}
\end{document}